\documentclass[12pt]{article}
\usepackage{float}
\usepackage{graphics,graphicx}
\usepackage{amsfonts}
\usepackage{amsmath}
\usepackage{natbib}
\newtheorem{theorem}{Theorem}

\newtheorem{corol}{Corollary}
\newtheorem{lemma}{Lemma}

\newcommand{\beq}{\begin{equation}}
\newcommand{\eeq}{\end{equation}}

\newcommand{\mx}{\mbox}
\newcommand{\ps}{{[0,\infty)}}
\newcommand{\iz}{{i,z}}
\newcommand{\jz}{{j,z}}
\newcommand{\ix}{{i,x}}
\newcommand{\jy}{{j,y}}

\newcommand{\wh}{\widehat}
\newcommand{\mc}{\mathcal}

\newcommand{\be}{\begin{eqnarray}}
\newcommand{\ee}{\end{eqnarray}}
\newcommand{\bea}{\begin{eqnarray*}}
\newcommand{\eea}{\end{eqnarray*}}
\newcommand{\bi}{\begin{itemize}}
\newcommand{\ei}{\end{itemize}}
\newcommand{\ben}{\begin{enumerate}}
\newcommand{\een}{\end{enumerate}}
\newcommand{\bay}{\begin{array}}
\newcommand{\eay}{\end{array}}
\newcommand{\bsl}{\begin{slide}}
\newcommand{\esl}{\end{slide}}
\newcommand{\bcen}{\begin{center}}
\newcommand{\ecen}{\end{center}}

\newcommand{\bs}{\boldsymbol}

\begin{document}

\title{Nonparametric Covariate Adjustment for Receiver Operating Characteristic Curves}

\author{
{\bf
Fang Yao }\\
Department of Statistics, University of Toronto, Canada\\ 
and\\
{\bf Radu V. Craiu}\\
Department of Statistics, University of Toronto, Canada\\
and\\
{\bf Benjamin Reiser }\\
Department of Statistics, University of Haifa, Israel}

\date{July 2007}

\maketitle

\bibliographystyle{Chicago}

\begin{abstract}

The accuracy of a  diagnostic test is typically characterised using
the receiver operating characteristic (ROC) curve. Summarising
indexes such as the area under the ROC curve (AUC)  are used to
compare different tests as well as to measure the difference between
two populations. Often additional information is available
 on some of the covariates which are known to influence the accuracy of such measures.
 We propose nonparametric methods for
 covariate adjustment of the AUC.  Models with normal errors
 and non-normal errors are discussed and analysed separately.
 Nonparametric regression is used for estimating mean and variance functions
 in both scenarios. In the general noise case we propose  a covariate-adjusted Mann-Whitney estimator for AUC estimation which effectively uses available data to  construct {\it working samples} at any covariate value of interest and is computationally efficient for implementation.
 This provides a generalisation of the Mann-Whitney approach for comparing two populations by taking covariate effects into account.
 We derive asymptotic properties for the
AUC estimators in both settings, including asymptotic normality, optimal strong
uniform convergence rates and MSE consistency.
The usefulness of the proposed methods is  demonstrated through simulated and real data examples.
\end{abstract}

\bigskip

{\bf Keywords:} {\em Area Under Curve, Asymptotics, Covariate Adjustment,   Mann-Whitney, Nonparametric, Smoothing, Uniform Convergence}

%\centerline{\large  ABSTRACT}

%\vspace{0.15in}

\section{Introduction}

The receiver operating characteristic (ROC) curve is a commonly used
tool for summarizing the accuracy of a test with binary results. The
sensitivity, or true positive rate, of a binary test is the
probability that a truly diseased subject is diagnosed as diseased.
The specificity, which is also equal to one minus false positive
rate, is defined as the probability that a healthy subject produces
a negative test. Suppose that the result of a test is a random
variable $Y$; depending on whether $Y < c$ or $Y \ge c$ the test
result is considered negative or positive, respectively. If the
distribution of $Y$ is continuous, each value of the threshold $c$
will correspond to different sensitivity and specificity values.
%Hanley (1989), among many others, argued that instead of summarizing
%the performance of the test for a particular $c$, it is preferable
%to sum up the properties of the test using the ROC curve, that is
%the plot of sensitivity against one minus specificity when $c$ is
%varied over the whole range. 
In general the ROC curve summarizes how
well two populations can be separated by a specified variable.
Frequently a number of tests (a.k.a. markers or classifiers) are
performed on each individual subject.  A global univariate summary
of the corresponding ROC curve is used to determine which classifier
is more accurate. A number of such summaries are available but the
most commonly used one is the {\it area under the ROC curve (AUC)}.
%and the Youden Index (YI) (Youden, 1950).
The AUC can be interpreted as the
probability that a randomly chosen diseased subject will have a
marker value greater than that of a randomly chosen nondiseased
subject and can be used as an alternative measure of difference between two 
populations (e.g. Zhou et al., 2002). Its range of application extends from medical 
applications to reliability theory (Reiser and Guttman, 1986).

% Wolfe and Hogg (1971) have proposed this probability as a
%measure of the difference between two populations and argued that
%this is often more meaningful than examining mean differences. Hauck
%et al. (2000) have emphasized its use as a measure of treatment
%effects in clinical trials. In addition, this probability naturally
%arises in reliability theory (Reiser and Guttman, 1986).

The presence of ROC curves has become ubiquitous in medical studies
(Metz, 1989; Hsiao et al., 1989; Aoki et al., 1997; Otto et al., 1998;
Stover et al, 1996; Zhou et al., 2002), its usage being spurred
by the now classic text of Swets and Pickett (1982). Parametric and
nonparametric methods for estimating individual ROC curves are
available as well as methods that do not assume independent
observations (Begg, 1991; Delong et al., 1988; Molodianovitch et al.,
2006; Pepe, 2003).

In a large number of situations,  additional information is available in the form of covariates which are known to influence the accuracy of the test. Only recently, statistical methods have
been devised to incorporate such information in the ROC-based
analysis.  Some of the earlier  methods have been produced by 
 Thompson and Zucchini (1989), Obuchowski (1995),  Tosteson and Begg (1988)
and Toledano and Gatsonis (1995).   Pepe (1997)
formulated a general regression framework to model the dependence of
the ROC curve directly on the covariates. Pepe (2000)  and Dodd and Pepe (2003)
propose semiparametric approaches to model the ROC  and AUC  directly using
 generalized linear models. Cai and Pepe (2002) extend the parametric ROC regression model by allowing an arbitrary nonparametric baseline function. Cai (2004) finds a more efficient   estimator in the semiparametric setting.
% Pepe (1998)
%summarized three approaches used for covariate
 %adjustment in ROC-based analyses.
Brumback et al. (2006) used an alternative procedure by applying a
generalized regression framework directly to  the AUC in order to
adjust the Mann-Whitney test for covariates. However, this approach loses the
connection with the threshold value, does not allow the prediction
of the sensitivity and specificity at a given threshold conditional
on covariates nor does it model covariate effects on the individual
marker values. Consequently we prefer to directly model the
covariate effects on the marker values and through this modeling
process obtain the analyses of interest.

The methods proposed in this paper fall within the first category of methods
described in Pepe (1998).  We  propose a nonparametric 
approach to adjust for covariates  the computation of AUC and other ROC-related quantities of interest.  The main motivation for our 
method is the  robustness to model mis-specification which may beset a parametric adjustment.
We thus generalize in two
ways the approaches of Faraggi (2003) and Schisterman et al.  (2006)
who use normal regression models to adjust the index AUC for
covariates. 
%Their approach is based on Guttman et al. (1988) who
%discussed covariate adjustment in a reliability context. 
We describe the regression model, distinguishing between the normal
noise assumption and the general noise assumption. In a first
extension of previous work, we estimate the mean and variance
functions using nonparametric regression techniques, more
specifically, local polynomial regression instead of parametric linear
models.  Our main contribution leading to the second extension is to construct 
a covariate-adjusted Mann-Whitney estimator (CAMWE) in the general noise case, which
relies on  {\it working samples} created at any possible covariate value $Z=z$ of
interest for the estimation of AUC. Such working samples have, for any
$Z=z$, the same size as the original sample and can be used to estimate a 
number of covariate-adjusted characteristics of the ROC curve. 
In practice the computation is kept minimal by utilizing the estimated mean and
variance functions for all $Z=z$ of interest. We recommend bootstrapping in order to
obtain confidence intervals for the covariate-adjusted AUC. Although we focus on covariate-adjusted AUC
estimation, the proposed methods can be
readily extended to other measures related to ROC curves, e.g., the
covariate-adjusted specificity, sensitivity and Youden Index (Youden, 1950).

A theoretical investigation provides  asymptotic results for both
the normal noise and general noise models. The asymptotic normality
and optimal strong uniform convergence rates for the
covariate-adjusted AUC estimators for normal noise are established. 
For the general noise distribution we first derive
asymptotic normality of the ``hypothetical'' CAMWE and then
characterize the asymptotic behavior of the Mean Squared Error (MSE)
of the CAWME.   We performed simulations  under a number of scenarios 
to demonstrate the effectiveness and robustness of the proposed estimators as well as the validity of the Bootstrap scheme for confidence band construction.

\section{ Model and Estimation}

\subsection{ Regression Model}

To motivate our proposal, we first note that parametric methods are used mainly for simple interpretation but may mis-specify the correct model forms, while nonparametric models provide an alternative solution and are more robust and data-adaptive. 
We attempt to achieve the robustness from two perspectives. First, we do not assume any parametric forms for the mean and variance functions  
of the test response variables, $X$ for nondiseased individuals $X$
and $Y$ for diseased individuals. Although we refer to ``diseased'' and ``nondiseased'' groups, the above framework applies to any two populations of interest. 
We utilize nonparametric regression
models
 \beq \label{reg0} X|Z=f(Z)+\sqrt{v_1(Z)} \,\epsilon_{1}, \eeq
\beq \label{reg1} Y|Z=g(Z)+\sqrt{v_2(Z)} \,\epsilon_{2}, \eeq where
$Z$ denotes the covariate, the standardized errors $\epsilon_1$ and
$\epsilon_2$ are independent of each other with zero mean and unit
variance, and the variance functions $0<v_1(z)<\infty$ and $0 < v_2(z)
<\infty$ for all $z\in \Re$.  Note that the errors here can depend
heteroscedastically on the covariate $Z$ through $v_1$ and $v_2$.  
Second, we do not assume specific distributions for the noises in order to guard against mis-speciffication of error distributions. Denote the conditional cumulative distribution functions (c.d.f.) of $X$ and $Y$ given $Z$ by $F(\cdot |Z)$ and $G(\cdot|Z)$, and c.d.f.s of $\epsilon_1$ and $\epsilon_2$ by $F^\ast(\cdot)$ and $G^\ast(\cdot)$.  Here we assume $F^\ast$ and $G^\ast$ do not depend on $Z$, i.e., the dependence of $X$ and $Y$ on $Z$ are expressed only through $f$, $g$, $v_1$ and $v_2$, which is equivalent to  a location-scale model. It is worth mentioning that, if the response variable is appropriately chosen at $Z=z$, then marker values of the diseased sample should be greater than that of the nondiseased sample on average. This is equivalent to $P(Y>X|Z=z)>0.5$, an assumption implicitly made for the remaining of the paper. If the baseline distributions $F^\ast$ and $G^\ast$ are symmetric about $0$, it implies the assumption $g(z)>f(z)$. In practice, we can simply constrain all the AUC estimators to be greater than 0.5. This would not affect any subsequent development due to the consistency of the unrestricted estimators as presented in Section 3. For notational convenience, we use the unrestricted forms throughout the paper.

This extends the first type of models discussed by Pepe (1998), where linear forms were assumed for $f$ and $g$ with variances not depending on the covariate $Z$, i.e., $g(z)=\alpha_0+\alpha_1+(\alpha_2+\alpha_3) z$, $f(z)=\alpha_0+\alpha_2 z$, $v_1(z)=v_1$ and $v_2(z)=v_2$. It is also noticed that we do not require the same baseline distributions of the standardized error $\epsilon_1$ and $\epsilon_2$ in contrast to Pepe (1998). Moreover, when the noise is not normally distributed, we shall propose a new estimator for the area under the ROC curve that extends the Mann-Whitney estimator for covariate-adjustment  by using standardized residuals via the so-called {\it working samples}.

\subsection{Estimation under Normal Noise Assumption}

Let $A(z)$ be the area under the ROC curve with the covariate adjustment $Z=z$. From models (\ref{reg0}) and
(\ref{reg1}), when the errors $\epsilon_1$ and $\epsilon_2$ are normally distributed, i.e., $F^\ast=G^\ast=\Phi$, where $\Phi(\cdot)$ is the
c.d.f. of the standard normal, it is straightforward to
derive the following explicit expression:
\beq \label{auc-n}
A_N(z)=P(Y>X|Z=z)=\Phi\left\{\frac{g(z)-f(z)}{\sqrt{v_1(z)+v_2(z)}}\right\},
\eeq
where the subscript ``$_N$'' stands for the normal assumption.
One can also obtain closed forms of the sensitivity $q_N(z)$ and specificity $p_N(z)$
for $Z=z$,
\beq \label{pq-n}
q_N(z)=\Phi\left\{\frac{g(z)-c}{\sqrt{v_2(z)}}\right\}, \hspace{0.3in}
p_N(z)=\Phi\left\{\frac{c-f(z)}{\sqrt{v_1(z)}}\right\}, \eeq for a given
threshold $c$.  The ROC curve for the covariate $Z=z$ is the
plot of $q(z)$ versus $1-p(z)$ for all possible values of $c$, and
this can be explicitly written as \beq \label{roc-n}
q_N(z)=\Phi\left[\frac{g(z)-f(z)+\sqrt{v_1(z)}\Phi^{-1}\{1-p(z)\}}{\sqrt{v_2(z)}}\right].
\eeq
The unknown functions $f, g, v_1, v_2$,
are estimated by using nonparametric regression methods as addressed in Section 3.1,
providing a ``nonparametric adjustment'' as discussed in Section 1.

\vspace{0.1in}

\subsection{ Estimation under General Noise Assumption}

The assumption of normal noise above simplifies the calculations of
the AUC via (\ref{auc-n}) but is not always supported by the data. In
addition, the normality assumption hampers the full generality one
expects from a nonparametric model. We propose here a fully
nonparametric yet simple estimator of the AUC with covariate
adjustment, $A(z)=P(Y>X|Z=z)$, for a general noise distribution.

The proposed estimator is motivated by the classical Mann-Whitney statistic,
which is formulated for two samples $\{x_1, \ldots, x_m\}$ and
$\{y_1, \ldots, y_n\}$ as 
\beq \label{mw}
M_{m,n}=\frac{1}{mn}\sum_{i=1}^m\sum_{j=1}^n 1_\ps(y_j-x_i),
\eeq
where $1_\ps(x)=1$ if $x\geq 0$ and $1_\ps(x)=0$ otherwise.
The data obtained from nondiseased and diseased samples consist of
$\{(z_{i,x}, x_i): i=1, \ldots, m\}$ and $\{(z_{j,y}, y_j): j=1, \ldots, n\}$,
where $z_{i,x}$ is the observed covariate value in the nondiseased sample and
$z_{j,y}$ in the diseased sample. It should be noticed that the markers $X$ and
$Y$ are evaluated at possibly different values of the covariate $Z$, and
we are often interested in estimating $A(z)$ even for z-values which were not
measured in either group or both.
To estimate $A(z)$ at  $Z=z$, one possibility is  to include the
marker values $x_i$ and $y_j$ that  fall into neighborhoods
of $z$ with appropriate weight functions. This consideration naturally leads to a bivariate kernel estimator that is fully nonparametric,
\beq \label{auc-l}
\widehat{A}_K(z)=\frac{\sum_{i=1}^m \sum_{j=1}^n 1_\ps(y_j-x_i) K_{h_x}(z_{i,x}-z)K_{h_y}(z_{j,y}-z)}{\sum_{i=1}^m \sum_{j=1}^n K_{h_x}(z_{i,x}-z)K_{h_y}(z_{j,y}-z)},
\eeq
where  $h_x$ and $h_y$ are  bandwidths, $K_h(\cdot)=(1/h)K(\cdot/h)$ when  $K(\cdot)$ is  a symmetric kernel density.
However, $\widehat{A}_K$, does not efficiently use the available data due to
the restriction on the local windows, nor do the regression models (\ref{reg0}) and (\ref{reg1}) play any role here. Note that $\widehat{A}_K$ is obtained by smoothing the binary variables  $1_\ps(y_j-x_i)$ corresponding to covariate observations  $(z_{i,x},z_{j,y})\in [z-h_x, z+h_x]\times [z-h_y, z+h_y]$ that are not necessarily located on the diagonal (in fact, $\{z_{i,x}\}$ and $\{z_{j,y}\}$ may have no overlap).  It is unclear how to choose the 
bandwidths  $h_x$ and $h_y$ which are critical to the kernel regression estimation, as the standard cross-validation procedure does not apply due to the absence of the observed $(z_{i,x},z_{j,y},1_{[0, \infty)}(y_j-x_i))$ on the diagonal of the bivariate covariate surface. 
More discussion and comparisons concerning $\widehat{A}_K(z)$ will be presented in  simulations in Section 4.

Based on the above considerations, we propose a different nonparametric estimator
of $A(z)$ which utilizes the entire collection
of data available and  the regression models (\ref{reg0}) and
(\ref{reg1}). First, suppose that we can observe all the
standardized residuals, $ i=1, \ldots, m$, $j=1,\ldots,n$, \beq
\label{sr} \epsilon_\ix={x_i-f(z_{i,x})\over \sqrt{v_1(z_{i,x})}},
\hspace{0.3in} \epsilon_\jy={y_j-g(z_{j,y})\over
\sqrt{v_2(z_{j,y})}}. \eeq Recall that the distributions of
$\epsilon_1$ and $\epsilon_2$ do not depend on $Z$, implying that
$\epsilon_{1,i}$ are independently and identically distributed
(i.i.d.) with the c.d.f. $F^\ast$ for $i=1, \ldots, m$, and
$\epsilon_{2,j}$ are i.i.d. with the c.d.f. $G^\ast$ for $j=1,
\ldots, n$. In Pepe (1998) these  standardized residuals can be used to obtain the
 empirical distributions of $\epsilon_1$ and $\epsilon_2$.
In a similar sprit, we propose a different way to utilize these residuals to 
construct {\it working samples} $\{x_{i,z}, \ldots, x_{m, z}\}$ and $\{y_{1,z}, \ldots, y_{n,z}\}$  as if they were all observed at $Z=z$, \beq \label{xy-z}
x_\iz=f(z)+\sqrt{v_1(z)} \epsilon_\ix, \hspace{0.3in}
y_\jz=g(z)+\sqrt{v_2(z)} \epsilon_\jy. \eeq Then it is intuitive to
use the proposed Covariate-Adjusted Mann-Whitney Estimator
(CAMWE) for $A(z)$, \beq \label{auc-m}
A_{M}(z)=\frac{1}{mn}\sum_{i=1}^m \sum_{j=1}^n 1_\ps(y_\jz-x_\iz).
\eeq 
This is a natural extension of  the Mann-Whitney estimator since in the case 
of no covariate effect   $f$, $g$, $v_1$, $v_2$ are constant in $z$ and
(\ref{auc-m}) becomes the traditional Mann-Whitney statistic. For
practical implementation, after obtaining nonparametric estimates of
$f, g, v_1$ and $v_2$, we do not have to
choose other tuning parameters for each covariate value $Z=z$, while
(\ref{auc-l}) requires retuning. Analogously we can calculate the
sensitivity and specificity from the working samples for $Z=z$, \beq
\label{pq-m} q_M(z)=\frac{1}{n}\sum_{j=1}^n 1_\ps(y_\jz \geq c),
\hspace{0.3in} p_M(z)=\frac{1}{m}\sum_{i=1}^m 1_\ps(x_\iz \leq c),
\eeq for a given threshold $c$. The ROC curves for $Z=z$ can be
obtained by plotting $q_M(z)$ versus $1-p_M(z)$ for all possible
values of $c$.
%, which is
%nondecreasing along $x$-axis but does not necessarily have one-to-one correspondence
%as that under normal  noise assumption.

{\it Remark 1.} Note that the central idea is to
construct the {\it working sample} $\{x_{i,z}, \ldots, x_{m, z}\}$ and
$\{y_{1,z}, \ldots, y_{n,z}\}$ for each $Z=z$. The entire conditional ROC curve, given
the  covariate value $Z=z$, can be obtained from (\ref{pq-m}). One can estimate any index of interest at
$Z=z$ using this working sample. For instance, the Youden Index (YI)
(Youden, 1950) can be calculated  by
$\mbox{YI}_M(z)=p_M(z)+q_M(z)-1$, where $p_M(z)$ and $q_M(z)$ are
defined by (\ref{pq-m}), and its optimal threshold given $Z=z$ can be
found via a numerical search.

{\it Remark 2.} In principle, the proposed approach can be
extended to the case of multiple covariates using different
strategies. A natural consideration is to use multivariate
nonparametric smoothing techniques that require extensive
computation. An alternative is to use additive frameworks for mean and
variance structures respectively, then construct the working sample in
a similar spirit for each set of covariate values of interest.

\subsection{Implementation via Nonparametric Regression}

%Although standard parametric models will find features in the data
%which have been already incorporated {\it a priori}, these models
%may not be adequate if the ``true'' patterns of $f$, $g$ $v_1$ and
%$v_2$ are not well defined and do not fall into a preconceived class
%of functions.  In such situations an analysis through nonparametric
%regression is advisable, since it  is a more flexible and
%data-adaptive way to characterise the functional relationship.

We exploit the local polynomial regression models for estimating the
functions $f$ and $g$.  Let $K(\cdot)$ be a compactly-supported
symmetric kernel density function with a finite variance,
$h_1=h_1(m)$ a sequence of bandwidths used to estimate $f$, and
$h_2=h_2(n)$ a sequence of  bandwidths for $g$. Let $p$ be the order
of local polynomial fit, e.g., $p=0$ and $p=1$ correspond to local
constant and local linear fits, respectively. An odd order fit is
often suggested (Fan and Gijbels, 1996) for both theoretical and
practical considerations. In particular, for estimating the
regression function itself, a common choice is the local linear fit
with $p=1$. Denote the resulting $p$th order local polynomial
estimators of $f(z)$ and $g(z)$ by $\hat{f}(z)$ and $\hat{g}(z)$.
Next, the variance functions $v_1(z)$ and $v_2(z)$ for
heteroscedastic errors  are estimated by fitting local polynomial
regression to the squared residuals, $v_{i,x}$ and $v_{j,y}$, $i=1,
\ldots, m, j=1, \ldots, n$, \beq \label{vo}
v_{i,x}=\{x_i-\hat{f}(z_{i,x})\}^2, \hspace{0.3in}
v_{j,y}=\{y_j-\hat{g}(z_{j,y})\}^2, \eeq with bandwidths
$b_1=b_1(m)$ and $b_2=b_2(n)$. The detailed formulas of the
aforementioned local polynomial estimators are given in Appendix
1. In the case of homoscedastic errors, $v_1(z)\equiv v_1$ and
$v_2(z)\equiv v_2$, it is easy to obtain root-n consistent
estimators (Hall and Marron, 1990; Hall et al., 1990).
%In practice, we adapt an efficient and
%asymptotically optimal difference-based estimation approach proposed
%by Hall et al. (1990) in homoscedastic regression models.
The theoretical properties in Section 4 are still valid with slight modifications. 
In practice, the bandwidths
$h_1$, $h_2$, $b_1$ and $b_2$ are chosen  by the standard technique of
leave-one-out cross-validation for estimating the mean and variance functions, while other existing techniques can certainly be applied.  %Although these selected bandwidths may be not optimal (see the Remark below), the simulation still provide some empirical support of the effectiveness (Figure \ref{simu-bc}).
Such bandwidths usually fulfill the assumptions needed for theoretical developments in Section 3 for sufficiently large sample sizes.
Substituting the local polynomial estimators $\hat{f}(z)$,
$\hat{g}(z)$, $\hat{v}_1(z)$ and $\hat{v}_2(z)$ for these unknown
quantities in formulae (\ref{auc-n})-(\ref{roc-n}), (\ref{auc-m}) and
(\ref{pq-m}) provides the point estimators $\wh{A}_N(z)$,
$\hat{p}_N(z)$, $\hat{q}_N(z)$, $\wh{A}_M(z)$ $\hat{p}_M(z)$ and
$\hat{q}_M(z)$ for covariate $Z=z$.

To evaluate confidence limits and variances for AUC under normal noise, the existing formulation (Guttman {\it et al.}, 1988; Faraggi, 2000, among others) are no
longer valid due to nonparametric regression. In principle we can derive the approximate variance for AUC under normal noise, based on the asymptotic normality of the local polynomial estimators (Fan and Gijbels, 1996) using the Cram\'er-Wold
device. However, due to the complicated asymptotic expressions with unknown functionals and their derivatives, the evaluation of such asymptotic
quantities will require extensive pilot smoothing and further approximations. This might
deteriorate the accuracy and not be worth
further pursuing.  Thus we choose to obtain confidence
limits and variance estimates for AUC via ``bootstrapping the original data'' as proposed by Efron and Tibshirani (1993). We do
not repeat the procedure here for conciseness. While this approach can
be justified in  normal noise case due to the limiting distributions in Theorem
\ref{thm1}, it may not be the case under the general noise for which the asymptotic normality of the CAMWE $\widehat{A}_M(z)$ is unknown at this moment. Nevertheless, the simulation performed in Section 4.1 offers empirical support to this bootstrap procedure for the general noise case.

 {\it Remark.}\quad Jointly choosing four bandwidths simultaneously aiming at the AUC estimator is prohibitively expensive, even impossible with available computing resources. Even if the computation load were not an issue, we would have no suitable criterion to perform the joint optimization for two reasons. First, if one bases the criterion on asymptotic bias and variance, these quantities involve unknown functionals and their derivatives and are too complicated for practical use. It should also be noticed that such asymptotic expressions are established only for the normal noise case. Second, if one attempts cross-validation directly for $A(z)$, there are no observed values of AUC at $Z=z$ available, which is a similar issue as the one discussed for $\widehat{A}_K$ in Section 2.3.

\section{ Theoretical Properties}
In this section we present the asymptotic theory developed for the
nonparametric estimators of the AUC with covariate adjustment for
$Z=z$ under both normal and general noise assumptions. One can
easily extend these arguments to obtain the corresponding
asymptotic theory for the sensitivity $q(z)$ and specificity $p(z)$
with a given threshold value $c$. These are not presented here for
conciseness.

\subsection{ Asymptotic Properties  under Normal Noise}
We begin with the asymptotic normality of the estimated AUC under
the normal noise assumption, where the target $A(z)$ is exactly
$A_N(z)$, i.e., $A(z)\equiv A_N(z)$. Let $\theta(z)$ be the density
function of the covariate $Z$ that is treated as a random variable.
Denote by $N(z)$ a neighborhood of $z$. Assume that, for a given
value $z$ of $Z$, \bi
\item[(A1)] $\theta(z)>0$ and $\theta(\cdot)$ is continuous in $N(z)$.
\ei Put $\eta_1(z)=E(\epsilon_1^3|Z=z)$,
$\eta_2(z)=E(\epsilon_2^3|Z=z)$,
$\kappa_1(z)=\mbox{Var}(\epsilon_1^2|Z=z)$ and
$\kappa_2(z)=\mbox{Var}(\epsilon_2^2|Z=z)$. Assume that, for a given
$z$, \bi
\item[(A2)] $v_1(z)>0$, $f^{(p+1)}(\cdot), v_1^{(p+1)}(\cdot),
\eta_1(\cdot)$ and $\kappa_1(\cdot)$ are continuous in $N(z)$.  \ei
Recall that $h_1=h_1(m)$, $b_1=b_1(m)$, $h_2=h_2(n)$ and $b_2=b_2(n)$
are the sequences of bandwidths for estimating $f(z)$, $v_1(z)$,
$g(z)$ and $v_2(z)$. One can see that, if the bandwidths $h_1$ and
$b_1$ are chosen optimally for estimating $f(z)$ and $v_1(z)$, then
$h_1$ and $b_1$ will be of the same order in terms of the sample size
$m$. Thus we assume the following, as $m\rightarrow \infty$, \bi
\item[(A3)] $h_1\rightarrow 0$, $mh_1\rightarrow \infty$,
$m h_1^{2p+3}\rightarrow d_1^2$ for some $d_1>0$, $b_1/h_1
\rightarrow \rho_1$ for some $0<\rho_1<\infty$.  \ei

Analogously, for the estimation of $g(z)$ and $v_2(z)$, we assume
that, for a given $z$, \bi
\item[(A4)] $v_2(z)>0$, $g^{(p+1)}(\cdot), v_2^{(p+1)}(\cdot),
\eta_2(\cdot)$ and $\kappa_2(\cdot)$ are continuous in $N(z)$;
\item[(A5)] $h_2\rightarrow 0$, $n h_2\rightarrow \infty$,
$n h_2^{2p+3}\rightarrow d_2^2$ for some $d_2>0$, and $b_2/h_2
\rightarrow \rho_2$ for some $0<\rho_2<\infty$.  \ei

Here we consider the odd order $p$ of local polynomial estimators
for $f$, $v_1$, $g$ and $v_2$ as argued in Section 2.4. The same
order $p$ is used mainly for notational convenience, while we
certainly can choose different orders in practice.  With slight
modifications, the results can be easily adapted to possibly
different orders as well as the case of even $p$.
For the symmetric kernel
density $K(\cdot)$ we assume that the j-th moment
$\mu_j(K)=\int u^jK(u)du$ exists for all integer $j\ge 0$.
 \bi
\item[(A6)] $R(K)=\int K^2(u)< \infty$, $\mu_2(K)>0$.
\ei For convenience, we introduce the notion of the order of a
kernel function. We say $K_0$ is an $\ell$th order kernel function,
provided that $\mu_0(K_0)=1$, $\mu_j(K_0)=0$ for $j=1,\ldots,
\ell-1$ and $\mu_\ell(K_0)\neq 0$.  It is obvious that $K(\cdot)$ is
a 2nd order kernel. Let the $(p+1) \times (p+1)$ matrix
$S_p=\{\mu_{j+l}(K)\}_{0\leq j, l\leq p}$, ${\bs e}_k$ be the
$(p+1)\times 1$ vector with the $k$th element equal to 1 and 0
elsewhere, and \be \label{equivK} K^\ast(u)={\bs e}_1^T
S_p^{-1}(1,u, \ldots, u^p)^T K(u), \ee which is often referred to as
the equivalent kernel. One can verify that $K^\ast(\cdot)$ is a
$(p+1)$th order kernel when $p$ is odd.  Also denote $R(K^\ast,
\rho)=\int K^\ast(u)K^\ast(u/\rho)du$ for any $0<\rho<\infty$.

Lemma 1 in Appendix 2 provides the joint asymptotic distributions of
the local polynomial estimators of $\{f(z), v_1(z)\}^T$ and
$\{g(z), v_2(z)\}^T$, which is the basis for deriving the
asymptotic distributions of $\wh{A}_N(z)$.  The difficulty in the proof of Lemma 1 is
to deal with the dependence between the mean and variance estimators,
while $\{\hat{f}(z), \hat{v}_1(z)\}^T$ and $\{\hat{g}(z), \hat{v}_2(z)\}^T$ are
independent, see Appendix 2 for details.
Based on Lemma 1, we
exploit the Cram\'{e}r-Wold device to obtain the asymptotic
distribution of $\wh{A}_N(z)$ as follows.

\begin{theorem} \label{thm1}
Under the assumptions (A1)-(A6) for a given $z$,
\bi
\item if $n/m\rightarrow \infty$,
$\sqrt{mh_1}\{\wh{A}_N(z)-A_N(z)\}\stackrel{D}{\longrightarrow}N\{B_{1}(z),
V_{1}(z)\}$, where $\phi(u)=(2\pi)^{-1/2}e^{-u^2/2}$,
$\delta(z)=\{g(z)-f(z)\}/\sqrt{v_1(z)+v_2(z)}$, 
\be \label{thm1.1}
B_{1}(z) &=& -\frac{\phi\{\delta(z)\}\mu_{p+1}(K^\ast)d_1}{(p+1)!\sqrt{v_1(z)+v_2(z)}}\left[f^{(p+1)}(z)+
\frac{\{g(z)-f(z)\}v_1^{(p+1)}(z)\rho_1^{p+1}}{2\{v_1(z)+v_2(z)\}}\right],   \nonumber 
\\
\hspace{0.5in}V_{1}(z) & = &\frac{\phi^2\{\delta(z)\}}{\theta(z)\{v_1(z)+v_2(z)\}}\left [ R(K^\ast)v_1(z)   \right .   \\
 &+& \left.
 \frac{\{g(z)-f(z)\}R(K^\ast,\rho_1)\eta_1(z)}{ \{v_1(z)+v_2(z) \}\rho_1}  +
 \frac{\{g(z)-f(z)\}^2R(K^\ast)\kappa_1(z)}{4\{v_1(z)+v_2(z)\}^2\rho_1}\right], \nonumber
\ee

\item if $n/m\rightarrow 0$,
$\sqrt{nh_2}\{\wh{A}_N(z)-A_N(z)\}\stackrel{D}{\longrightarrow}N\{B_{2}(z),
V_{2}(z)\}$, where
 \be \label{thm1.2}
B_{2}(z)&=&\frac{\phi\{\delta(z)\}\mu_{p+1}(K^\ast)d_2}{(p+1)!\sqrt{v_1(z)+v_2(z)}}\left[g^{(p+1)}(z)-
\frac{\{g(z)-f(z)\}v_2^{(p+1)}(z)\rho_2^{p+1}}{2\{v_1(z)+v_2(z)\}}\right],\nonumber
\\
\hspace{0.5in} V_{2}(z)&=&\frac{\phi^2\{\delta(z)\}}{\theta(z)\{v_1(z)+v_2(z)\}}\left[R(K^\ast)v_2(z)  \right .  \\
&-&
\frac{\{g(z)-f(z)\}R(K^\ast,\rho_2)\eta_2(z)}{\{v_1(z)+v_2(z)\}\rho_2} \left.+\frac{\{g(z)-f(z)\}^2
R(K^\ast)\kappa_2(z)}{4\{v_1(z)+v_2(z)\}^2\rho_2}\right],\nonumber \ee

\item if $n/m\rightarrow \lambda$ for some $0< \lambda< \infty$,
$\sqrt{mh_1}\{\wh{A}_N(z)-A_N(z)\}\stackrel{D}{\longrightarrow}N\{B_{3}(z),
V_{3}(z)\}$, where \be \label{thm1.3}
B_{3}(z)=B_{1}(z)+\lambda^{-\frac{p+1}{2p+3}}B_{2}(z),\hspace{0.2in}
V_{3}(z)=V_{1}(z)+\lambda^{-\frac{2p+2}{2p+3}}V_{2}(z) \ee \ei
\end{theorem}

Besides the pointwise limiting distributions, we also establish
the optimal rates for strong uniform convergence of $\wh{A}_N$ 
in Theorem 2. Denote by $\mc{Z}$ the set of possible values of $Z$ (usually an interval on the real line). Additional assumptions below are needed for the uniform convergence
results,
\ben
\item[(A7.1)] $E(|X|^s)<\infty, \sup_{z\in \mc{Z}}\int |x|^s p_{(Z, X)}(z, x)dx <\infty$
for some $s\geq 2$, where $p_{(Z, X)}$ is the joint density of $(Z, X)$.
\item[(A7.2)] $E(|Y|^s)<\infty, \sup_{z\in \mc{Z}}\int |y|^s p_{(Z, Y)}(z, y)dy
<\infty$ for some $s\geq 2$, where $p_{(Z, Y)}$ is
the joint density of $(Z, Y)$.
\een For the  proof of Theorem 2 we need to modify   (A1)-(A6) as follows. 
For convenience we impose conditions on the equivalent kernel $K^\ast$ (\ref{equivK}) instead of  the original kernel $K$.
\ben
\item[(A1$^\dag$)] $\theta(\cdot)>0$, and $\theta^{(p+1)(\cdot)}$
is bounded and continuous on $\mc{Z}$.
\item[(A2$^\dag$)]  On the domain $\mc{Z}$, $v_1(\cdot)>\delta_1$ for some
$\delta_1>0$ and is bounded, $f(\cdot)$ is bounded,
$f^{(p+1)}(\cdot)$, $v_1^{(p+1)}(\cdot)$, $\eta_1(\cdot)$ and $\kappa_1(\cdot)$ are bounded and continuous.
%\item[(A3$^\dag$.1)] $h_1\rightarrow 0$, $m^\rho h_1\rightarrow \infty$ for some $\rho<1-s^{-1}$,
%where $s$ satisfies (A7).
%\item[(A3$^\dag$.2)] $h_1\rightarrow 0$, $m^{2\rho-1} h_1 \rightarrow \infty$ for some $\rho<1-s^{-1}$,
%where $s$ satisfies (A7) and $s>2$.
%\item[(A3$^\dag$.3)] $\sum_m h_1^{\Delta_1}<\infty$ for some $\Delta_1>0$,
%$m^\rho_1 h_1\rightarrow \infty$ for some $\rho<1-s^{-1}$, where $s$ satisfies (A7).
\item[(A3$^\dag$)] $\sum_m h_1^{\Delta_1}<\infty$ for some $\Delta_1>0$,
$m^{2\rho_1-1} h_1\rightarrow \infty$ for some $\rho_1<1-s^{-1}$, where $s>2$
satisfies (A7.1).
\item[(A4$^\dag$)]  On the domain $\mc{Z}$,
$v_2(\cdot)>\delta_2$ for some $\delta_2>0$ and is bounded,
$g(\cdot)$ is bounded, $g^{(p+1)}(\cdot)$, $v_2^{(p+1)}(\cdot)$, $\eta_2(\cdot)$ and $\kappa_2(\cdot)$
are bounded and continuous.
\item[(A5$^\dag$)] $\sum_n h_2^{\Delta_2}<\infty$ for some $\Delta_2>0$,
$n^{2\rho_2-1} h_2\rightarrow \infty$ for some $\rho_2<1-s^{-1}$,
where $s>2$ satisfies (A7.2).
\item[(A6$^\dag$)] $K^\ast$ is uniform continuous, absolutely integrable
with respect to Lebesgue measure on $\Re$ and  of bounded variation,
$K^\ast(u)\rightarrow 0$ as $|u|\rightarrow \infty$, $\int \{|u\log(|u|)|\}^{1/2}
|d K^\ast(u)|<\infty$.
\een

Lemma 2 in Appendix 2 presents the strong uniform convergence rates
of the local polynomial estimators of the mean and  variance
functions.  Then the strong uniform convergence
rate of $\wh{A}_N$ is obtained immediately below, where a.s. is the
abbreviation of ``almost surely''.
\begin{theorem} \label{thm2}
Under the assumptions (A1$^\dag$)-(A6$^\dag$), (A7.1) and (A7.2), let
$\tau_m=h_1^{p+1}+\sqrt{\log(1/h_1)/(m h_1)}$ and
$\omega_n=h_2^{p+1}+ \sqrt{\log(1/h_2)/(n h_2)}$, then \beq
\label{thm2.1} \sup_{z \in
\mc{Z}}|\wh{A}_N(z)-A_N(z)|=O(\tau_m+\omega_n) \hspace{0.3in}
\mbox{a.s.}\eeq
\end{theorem}

\subsection{ Asymptotic Properties  under General Noise}

Now we turn to the asymptotic properties of the CAMWE $\wh{A}_M(z)$ of $A(z)$ under the general noise assumption. We first state the asymptotic normality of the ``hypothetical'' estimator $A_M(z)$ (\ref{auc-m}) that contains true
values of the unknown mean and variance functions, while our target
is $A(z)=P(Y>X|Z=z)$. Recall that $F^\ast$ and $G^\ast$ are the
c.d.f.s of standardized errors $\epsilon_1$ and $\epsilon_2$, and do
not depend on  the covariate $Z$. Define \beq \label{h-fn}
h_{1,0}(\epsilon_1; z)=G^\ast\left\{
\sqrt{\frac{v_1(z)}{v_2(z)}}\epsilon_1+\frac{f(z)-g(z)}{\sqrt{v_2(z)}}\right\}, \nonumber
\eeq
\beq
 h_{0, 1}(\epsilon_2; z)=F^\ast
\left\{\sqrt{\frac{v_2(z)}{v_1(z)}}\epsilon_2+
\frac{g(z)-f(z)}{\sqrt{v_1(z)}}\right\}. \nonumber \eeq Set 
$\xi_{1, 0}^2(z)=\mx{var}\{h_{1, 0}(\epsilon_1; z)\}$ and $\xi_{0,
1}^2(z)= \mx{var}(h_{0, 1}\{\epsilon_2; z)\}$.
\begin{theorem} \label{thm3}
For the regression models (\ref{reg0}) and (\ref{reg1}) and a given $z$,
\beq \label{thm3.1}
E\{A_M(z)\}=A(z), \hspace{0.3in}
\mx{var}\{A_M(z)\}=O\left(\frac{1}{m+n}\right).
\eeq
If $n/m\rightarrow \lambda$ for some $0 < \lambda <\infty$, $\xi^2_{1, 0}(z)>0$ and
$\xi^2_{0, 1}(z)>0$, then
\beq \label{thm3.2}
\sqrt{m+n}\{A_M(z)-A(z)\}\stackrel{D}{\longrightarrow}
N\left\{0, \frac{\xi_{1, 0}^2(z)}{\lambda^\ast}+\frac{\xi_{0, 1}^2(z)}{1-\lambda^\ast}\right\},
\eeq
where $\lambda^\ast=1/(1+\lambda)$.
\end{theorem}

In the next theorem we establish the MSE consistency of the
CAMWE $\wh{A}_M(z)$ for the
``hypothetical'' estimator $A_M(z)$ for a given covariate $Z=z$,
based on uniform consistency of the estimated mean and variance
functions. It is noticed in the proof that we actually do not need
the optimal strong uniform convergence rates stated in Lemma 2, as
these rates cannot be passed to $\wh{A}(z)$, while uniform consistency in probability is sufficient. Thus the regularity conditions (A3$^\dag$) and
(A5$^\dag$) can be relaxed to the following. \ben
\item[(A3$^\ast$)] $h_1\rightarrow 0$,
$m^{\rho_1} h_1\rightarrow \infty$ for some $\rho_1<1-s^{-1}$, where $s$ satisfies (A7.1).
\item[(A5$^\ast$)] $h_2 \rightarrow 0$,
$n^{\rho_2} h_2\rightarrow \infty$ for some $\rho_2<1-s^{-1}$,  where $s$ satisfies (A7.2).
\een
We also need the following additional assumptions,
\ben
\item[(A8)] $F^\ast(\cdot)$ and $G^\ast(\cdot)$ are continuous on their domains.
\een
\begin{theorem} \label{thm4}
Under (A8) and the assumptions for Theorem 2
with (A3$\dag$) and (A5$\dag$) replaced by (A3$^\ast$) and (A5$^\ast$),
for a given $z$,
\beq \label{thm4.1}
E[\{\wh{A}_M(z)-A_M(z)\}^2]\longrightarrow 0.
\eeq
\end{theorem}
We conclude this section with the following corollary that is a
direct consequence of Theorem \ref{thm3} and \ref{thm4}. Note
that the MSE discrepancy between estimated and true AUC at $Z=z$ is
dominated by the nonparametric rate in (\ref{thm4.1}) that is
usually slower the the parametric rate $(m+n)^{-1/2}$, although its
order of magnitude  is not obtainable, at least to our knowledge.
\begin{corol} \label{cor1}
Under (A8) and the assumptions for Theorem 2
with (A3$\dag$) and (A5$\dag$) replaced by (A3$^\ast$) and (A5$^\ast$),
for a given $z$,
\beq \label{cor1.1}
E[\{\wh{A}_M(z)-A(z)\}^2]\longrightarrow 0.
\eeq
\end{corol}

\section{ Simulations and Data Example}

\subsection{ Simulations}

The purpose of the simulations is to assess the performance of the
methods for estimating AUC in nonparametric regression settings. We
have not compared our method with parametric models since the two
approaches address different situations. If a parametric model is
correctly specified, its performance will be superior to a
nonparametric procedure; however, if there is no known parametric
model suitable for the data considered, one will have no choice but to
use the nonparametric tools available.

%There are infinitely many choices for the mean and variance
%functions so our study is certain to not cover all the
%scenarios of interest. However, we tried to incorporate in our
%functions patterns that are not uncommon, at least as far as our
%experience goes.

We consider three situations for illustration. In the first situation
the underlying models are, for non-diseased and diseased individuals
respectively, \be \label{simu-model1} x_i&=&6+1.5 z_{i, x} + 1.5
\sin(z_{i, x})+\sqrt{v_1(z_{i,x})}\ \epsilon_{i, x} \nonumber \\
y_j&=&6 + 1.5 z_{j, y} +1.5 \sin(z_{j, y}) + \sqrt{z_{j, y} -
0.5}+\sqrt{v_1(z_{i,x})}\ \epsilon_{j, y}, \ee where the errors
$\epsilon_{i, x}$ and $\epsilon_{j, y}$ are standard normal, the
conditional variance functions are $v_1(z)=0.3 + \Phi(2 z - 6)$ and
$v_2(z)=1.5 + \Phi(2 z - 6)$,  $i=1, \ldots, m$, $j=1, \ldots, n$. The covariates
$z_{i, x}$ and $z_{j,y}$ are independently generated from $U[1,5]$, and moderate sample sizes $n=m=40$ are used. The identical setting is used in the second situation,
except that the errors $\epsilon_{i, x}$ and $\epsilon_{j, y}$ are
generated from a Student-$t$ distribution with 3 degrees of freedom
 and rescaled to have zero mean
and unit variance.

The third situation, in which   the log-transformed responses
have normal errors $\epsilon_{i, x}^\ast$ and $\epsilon_{j, y}^\ast$,
i.e., the responses are generated from log-normal models, 
is designed to demonstrate the robustness of the
proposed CAMWE $\widehat{A}_M(z)$. Since a
log-transform often stabilizes the variability, we assume a constant
variance $\sigma^2$ on log-scale for both groups. Let $f_0(\cdot)$ and
$g_0(\cdot)$ be the mean functions on log-scale, while $f(\cdot)$ and
$g(\cdot)$ correspond to the original scale. From the properties of the
log-normal distribution, one has \bea \log\{f(z)\}=f_0(z)+\sigma^2/2,
&& \hspace{0in} v_1(z)=(e^{\sigma^2}-1) f^2(z)\\
\log\{g(z)\}=g_0(z)+\sigma^2/2, &&\hspace{0in} v_2(z)=(e^{\sigma^2}-1)
g^2(z).  \eea 
We choose $f(z)=1-0.5 z-0.25 \sin(\pi z)$ and $g(z)=1-0.5
z-0.25 \sin(\pi z)+1.5 \sqrt{z+0.5}$, $z \in [0, 1]$, and
$\sigma^2=1/3$. Then the models are completely determined and can be
written as \be \label{simu-model2} x_{i}=\exp\{f_0(z_{i,x})+ \sigma
\epsilon_{i, x}^\ast\}, \hspace{0.3in} y_{j}=\exp\{g_0(z_{j,y})+
\sigma \epsilon_{j, y}^\ast\}, \ee where the covariates $z_{i,x}$ and
$z_{j,y}$ are independently generated from $U[0,1]$, $\epsilon^\ast_{i,x}$ and $\epsilon^\ast_{j,y}$ are standard normal errors, $i=1, \ldots, m$,
$j=1, \ldots, n$.

With the generated data we  compared three estimators, $\wh{A}_N(z)$ with normal noise assumption, CAMWE $\wh{A}_M(z)$ with general noise assumption as well
as the kernel estimator $\wh{A}_K(z)$. For bandwidth choices, recall that joint selection aiming for $\wh{A}_N(z)$ and $\wh{A}_M(z)$ is not feasible and that
cross-validation fails for $\wh{A}_K(z)$. To make the comparisons possible,
for $\wh{A}_N(z)$ and $\wh{A}_M(z)$ we minimized the true integrated squared errors respectively, say $\int \{\hat{f}(z; h_1)-f(z)\}^2 dz$ to select $h_1$, and similarly for $h_2$, $b_1$ and $b_2$, while $\int \{\wh{A}_K(z; h_x, h_y)-A(z)\}^2 dz$ was minimized for choosing $h_x$ and $h_y$ in $\wh{A}_K(z)$. One can see that,  if one targets at $A(z)$, the bandwidths chosen for $\wh{A}_N(z)$ and $\wh{A}_M(z)$ may not be as ``optimal'' as those for $\wh{A}_K(z)$. However, it is demonstrated below that even in such a disadvantageous situation, the proposed estimators, especially $\wh{A}_M(z)$, are still preferable. We used the sample sizes of $n=m=40$ and $n=m=100$, while all the estimates were improved  with increased sample sizes as expected. 
 All three AUC estimates are obtained by applying the estimation procedures to the simulated data $\{(z_{i, x}, x_i)\}_{i=1, \ldots, m}$ and $\{(z_{j, y},
y_j)\}_{j=1, \ldots, n}$ (on original scale throughout) in the aforementioned
three situations. Monte Carlo averages (calculated from 500 runs in
each case) of Mean Squared Errors at different values of $z$ are
presented in Figure 1.
We can see that, for the normal noise model the CAMWE $\wh{A}_M(z)$ and
normal estimator $\wh{A}_N(z)$ are comparable and both outperform the
kernel estimator $\wh{A}_K(z)$. Although $\wh{A}_K(z)$ improves upon
$\wh{A}_N(z)$ under the heavy-tailed Student-$t$ noise model, the
CAMWE $\wh{A}_M(z)$ is still the most effective. For the
log-normal model, when we apply these three
estimation procedures to the original responses, the CAMWE and kernel
estimators yield comparable results (CAMWE seems slightly better),
and both significantly improved upon the normal estimator. 

Now we examine the  empirical performance of the pointwise confidence bands and variance estimates obtained  by ``bootstrapping the data'' in the general noise case, i.e. when the CAMWE $\wh{A}_M(z)$ is used for estimation, we carried out an additional study. We used the same settings for the       
three models with normal, Student with  3 degrees of freedom and log-normal noises, respectively. The benchmark used for comparison
 is the 95\% pointwise confidence bands and variance estimates averaged from 500 Monte Carlo runs. In each Monte Carlo run,  $\widehat{A}_M(z)$ was obtained and
we bootstraped the data 1000 times to calculate 95\% bootstrap bands (defined between the 2.5th and the 97.5th percentiles) and bootstrap sample variance.  All the bandwidths involved in the estimation are selected respectively by         
leave-one-out cross-validation in smoothing steps. In the top panels of Figure \ref{simu-bc} we reported, for all three data-generating models with moderate sample sizes $n=m=40$, the comparisons between the Monte Carlo averages of  the bootstrap bands and the Monte Carlo bands.  In the bottom panels, similar comparisons were shown for the averaged bootstrap variance estimates of $\widehat{A}_M(z)$ against the Monte Carlo variances. 
From Figure \ref{simu-bc}, for the CAMWE         
$\widehat{A}_M(z)$, the averages of confidence bands obtained by ``bootstrapping the data'' approximate well the 95\% pointwise Monte Carlo bands. The same can be said about the averages of bootstrap variance estimates. This provides some empirical evidence for using the bootstrap confidence bands and variance estimates for the CAMWE $\widehat{A}_M(z)$ in the general noise case. 
For the normal noise model, we have done similar comparisons and  the results are almost identical to those obtained for $\widehat{A}_M(z)$ (thus not reported for brevity).
%Additionally, the agreement between the averages of $\widehat{A}_M(z)$ and the true $A(z)$ also supports the use of cross-validation for selecting bandwidths in respective smoothing step.

\subsection{ Real Data Example}

We consider the white onions data originally reported by Ratkowski
(1983) on the density-yield relationship of varieties of white Spanish
Onion grown in various regions of Australia. The data has been the
subject of a nonparametric analysis of covariance in Young and Bowman
(1995). One can see from Figure \ref{fig:wo}
%\ref{fig:wo}
that the relationship between the density and yield is non-linear for
the two regions considered here: Virginia and Purnong Landing.  A
question of interest is whether the two regions of origin for the
onions can be separated simply by looking at the  yield. Figure \ref{fig:wo} 
shows that the difference between yields depends on the
density which will be the covariate under consideration in our study.

If we apply directly the method of Faraggi (2003) to the data on the
original scale we observe a large discrepancy between the parametric
and nonparametric analyses, as illustrated by the top panel in Figure
\ref{fig:woauc}.  We also notice that bootstraping the data produces
wider 95\% confidence bands for large values of the density due to the sparseness and high variability. But even such confidence bands do not cover the parametric estimators of the AUC. We should note that due to the sparseness of observations with densities larger than 150 we focus on the covariate range $(0,150)$.
On the logarithmic scale, the relationship between yield and density
is more linear as can be seen from the bottom panel in Figure \ref{fig:wo}. In addition, the transformation seems to stabilize the variance so 
 it is not unexpected that he difference between the nonparametric approach and the
parametric one diminishes. 
% while the parametric estimate is not well covered by the 95\% Bootstrap confidence bands. 
We can also notice that, on both original and logarithmic scales, the estimates obtained under the normal assumption are more conservative indicating a smaller AUC for small densities.  This indicates that the normal assumptions
may not be  valid  for this dataset and that the nonparametric
approach is more suitable due to its robustness.

\section{ Conclusions}

We introduce nonparametric adjustment for covariate information in the
context of ROC analysis, more specifically for the AUC index.  The
essential idea in our proposal is that the conditional ROC curve
and all the indexes associated with
it (e.g. Youden Index (YI) and its optimal cutoff value)
 can be computed using the statistical  model and, subsequently, the reconstructed 
  {\it working sample}.
 The theoretical properties of the index estimators deserve further
investigation. The approach bears some similarity to the work on
nonparametric adjustment for covariates when estimating a treatment
effect as in Young and Bowman (1995) and Cantoni and de Luna (2006)
and advances in that field are likely to yield newer results for the
ROC covariate adjustment.  In contrast to their work we focus on a generalized 
Mann-Whitney  approach.  Our simulations demonstrate effectiveness
and robustness of the proposed method. While the discussion is
limited to the case of only one covariate,  the proposed approach 
can  be extended to multiple covariates in various ways (e.g.\., additive models).
It is expected that the computational load will significantly increase with each additional covariate added to the model. In principle one may consider reasonable parametric approximations suggested by nonparametric approaches that lead to simpler interpretations. For instance, one possibility is to use parametric models for the mean and variance functions following the nonparametrically estimated forms.  Similar strategy applies to approximating the empirical c.d.f. of the noise by parametric functions.

\section*{Appendix 1:  Local Polynomial Estimators} 

Recall that $\{(z_{i,x}, x_i)\}_{1\le i \le m}$ and $\{(z_{j,y}, y_j)\}_{1\le j\le n}$ are nondiseased and diseased samples.
The local polynomial regression estimator of $f(z)$ is obtained by
minimizing \beq \label{lp} \sum_{i=1}^m \{x_i-\sum_{k=0}^p \beta_k
(z_{i,x}-z)^k\}^2 K_{h_1}(z_{i,x}-z), \eeq where $h_1=h_1(m)$ is the
bandwidth controlling the amount of smoothing, and
$K_{h_1}(\cdot)=K(\cdot/h_1)/h_1$.  It is more convenient to work
with matrix notation. Denote the design matrix of (\ref{lp}) by
$Z_x$, \bea
Z_x=\left(\bay{cccc}1&(z_{1,x}-z)&\cdots& (z_{1,x}-z)^p\\
\vdots&\vdots&\quad&\vdots\\ 1&(z_{m,x}-z)&\cdots&
(z_{m,x}-z)^p\eay\right), \eea and put
$W_{x,h_1}=\mbox{diag}\{K_{h_1}(z_{i,x}-z): i=1, \ldots, m\}$ and
${\bs x}=(x_1, \ldots, x_m)^T$.  The local polynomial estimator is
then given by \be \label{lp-x} \hat{f}(z)={\bs e}_1^T (Z_x^T W_{x,
h_1} Z_x)^{-1} Z_x W_{x,h_1} {\bs x}. \ee
Analogously for the diseased sample $(z_{j,y}, y_j), j=1, \ldots, n$, the design matrix $Z_y$ and weight matrix $W_{y,h_2}$ are similarly defined, letting ${\bs y}=(y_1, \ldots, y)^T$,
then the local polynomial estimator for $g$ is
$ \label{lp-y} \hat{g}(z)={\bs
e}_1^T (Z_y^T W_{y,h_2} Z_y)^{-1} Z_y W_{y,h_2} {\bs y}.  
$

We next estimate the variance functions $v_1(z)$ and $v_2(z)$ for
heteroscedastic errors according to
models (\ref{reg0}) and (\ref{reg1}). The nonparametric estimators
$\hat{v}_1(z)$ and $\hat{v}_2(z)$ are obtained by fitting local
polynomial regression to the squared residuals, i.e., the variance
observations $v_{i,x}$ and $v_{j,y}$ as in (\ref{vo}).  Let $b_1=b_1(m)$ and $b_2=b_2(n)$ be the sequences of
bandwidths for $\hat{v}_1(z)$ and
$\hat{v}_2(z)$. Denote ${\bs v}_x=(v_{1,x}, \ldots, v_{m,x})^T$ and
${\bs v}_y=(v_{1,y}, \ldots, v_{n,y})^T$, we have 
\be \label{sigma-est} 
\hat{v}_1(z)={\bs e}_1^T (Z_x^T W_{x,b_1} Z_x)^{-1}
Z_x W_{x,b_1} {\bs v}_x,\hspace{0.2in} \hat{v}_2(z)={\bs e}_1^T (Z_y^T
W_{y,b_2} Z_y)^{-1} Z_y W_{y,b_2} {\bs v}_y,  \nonumber \ee where $Z_x$ and $Z_y$
are defined as the above, $W_{x,b_1}=\mbox{diag}\{K_{b_1}(z_{i,x}-z):
i=1, \ldots, m\}$ and $W_{y,b_2}=\mbox{diag}\{K_{b_2}(z_{j,y}-z): j=1,
\ldots, n\}$.

\section*{Appendix 2:  Auxiliary Results and Proofs}

\begin{lemma} \label{lem1}
If the assumptions (A1)-(A3), (A6) hold, and $m\rightarrow \infty$,
for a given $z$, \be \label{lem1.1} \sqrt{mh_1}\{\hat{f}(z)-f(z),
\hat{v}_1(z)-v_1(z)\}^T \stackrel{D}{\longrightarrow}N\{{\bs b}_1(z),
\Sigma_1(z)\}, \ee where ${\bs b}_1(z)=\{b_{11}(z), b_{12}(z)\}^T$ and
$\Sigma_1(z)=\{\sigma_{x,ij}(z)\}_{1\leq i, j\leq 2}$ with \bea
 b_{11}(z)=\frac{\mu_{p+1}(K^\ast)}{(p+1)!} d_1
f^{(p+1)}(z), \hspace{0.2in}
b_{12}(z)=\frac{\mu_{p+1}(K^\ast)}{(p+1)!} d_1 \rho_1^{p+1}
v_1^{(p+1)}(z), \eea \bea 
\sigma_{x,11}(z)=\frac{R(K^\ast)v_1(z)}{\theta(z)} , \ \
\sigma_{x,22}(z)=\frac{R(K^\ast)\kappa_1(z)}{\theta(z)\rho_1},\ \
\sigma_{x,12}(z)=\frac{R(K^\ast, \rho_1)\eta_1(z)}{\theta(z) \rho_1}.
\eea Analogously, if the assumptions (A1), (A4)-(A6) hold, and
$n\rightarrow \infty$, for a given $z$, \be \label{lem1.2}
\sqrt{nh_2}\{\hat{g}(z)-g(z), \hat{v}_2(z)-v_2(z)\}^T
\stackrel{D}{\longrightarrow}N\{{\bs b}_2(z), \Sigma_2(z)\}, \ee where
${\bs b}_2(z)=\{b_{21}(z), b_{22}(z)\}^T$ and
$\Sigma_2(z)=\{\sigma_{y,ij}(z)\}_{1\leq i, j\leq 2}$ with \bea
 b_{21}(z)=\frac{\mu_{p+1}(K^\ast)}{(p+1)!} d_2
g^{(p+1)}(z), \hspace{0.2in}
b_{22}(z)=\frac{\mu_{p+1}(K^\ast)}{(p+1)!} d_2 \rho_2^{p+1}
v_2^{(p+1)}(z), \eea \bea 
\sigma_{y,11}(z)=\frac{R(K^\ast)v_2(z)}{\theta(z)} , \ \
\sigma_{y,22}(z)=\frac{R(K^\ast)\kappa_2(z)}{\theta(z)\rho_2},\ \
\sigma_{y,12}(z)=\frac{R(K^\ast, \rho_2)\eta_2(z)}{\theta(z) \rho_2}.
\eea
\end{lemma}

\noindent {\it Proof of Lemma \ref{lem1}}.\ \ The asymptotic
normality of $\hat{f}(z)$ with the bias $b_{11}$ and the variance
$\sigma_{x, 11}$ is standard in local polynomial regression. Let
$v^\ast_{i,x}=\{x_i-f(z_{i,x})\}^2$, note that the input data
$v_{i,x}=\{x_i-\hat{f}(z_{i,x})\}^2=v^\ast_{i,x}+2\{x_i-f(z_{i,x}\}\{\hat{f}\{z_{i,x}-f(z_{i,x})\}+\{\hat{f}(z_{i,x})-f(z_{i,x})\}^2$.
Applying a local polynomial fit to $(z_{i,x},v_{i,x}), i=1, \ldots,
m$, one can see that the second term will result in a quantity of
the order $o_p(b_1^{p+1}+1/\sqrt{mb_1})$ and the third term will yield
$O_p\{h_1^{2(p+1)}+1/(mh_1)\}$. It is obvious that both quantities are
ignorable, compared to the local polynomial estimator $v_1^\ast(z)$
obtained by fitting $(z_{i,x},v^\ast_{i,x})$. Therefore the
estimators $\hat{v}_1(z)$ and $v_1^\ast(z)$ are asymptotically
equivalent with the same limit distribution.  Again we apply the
standard argument of local polynomial regression to obtain the
asymptotic normality of $\hat{v}_1(z)$ with the bias $b_{12}$ and
variance $\sigma_{x,22}$. To derive the covariance of the limit
distribution between $\hat{f}(z)$ and $\hat{v}_1(z)$, one can 
equivalently work with $\hat{f}(z)$ and $v^\ast(z)$. 
Using the equivalent kernel notation $K^\ast$, the limiting
covariance is identical to the following,  obtained by employing a Taylor expansion,
\bea
\mbox{cov}\{\tilde{f}(z)-f(z),
\tilde{v}_1(z)\}=\frac{1}{mh_1 \rho_1
\theta(z)}\{\int K^\ast(u)K^\ast(u/\rho_1)du \eta_1(z)+O(h)\}.
\eea
where 
\bea
\tilde{f}(z)=\frac{1}{mh_1\theta(z)}\sum_{i=1}^m
K^\ast(\frac{z_{i,x}-z}{h_1})x_i,\hspace{0.2in}
\tilde{v}(z)=\frac{1}{mb_1\theta(z)}\sum_{i=1}^m
K^\ast(\frac{z_{i,x}-z}{b_1})v_i^\ast.  \eea 
The same arguments can be applied to obtain the
joint asymptotic distribution in (\ref{lem1.2}).

{\it Proof of Theorem \ref{thm1}}.\ \  The Cram\'er-Wold device is
exploited to derive the asymptotic distributions of $\wh{A}_N(z)$ for
three possible cases, and the detailed proof is omitted for conciseness.

\begin{lemma} \label{lem2}
If the assumptions (A1$^\dag$)-(A3$^\dag$), (A6$^\dag$) and (A7.1) hold,
and $m \rightarrow \infty$,
\beq \label{lem2.1}
\sup_{z\in \mc{Z}}|\hat{f}(z)-f(z)|=O(\tau_m), \hspace{0.3in}
\sup_{z\in \mc{Z}}|\hat{v}_1(z)-v_1(z)|=O(\tau_m), \hspace{0.2in} w.p.1.,
\eeq
and If the assumptions (A1$^\dag$), (A4$^\dag$)-(A6$^\dag$) and (A7.2) hold,
and $n \rightarrow \infty$,
\beq \label{lem2.2}
\sup_{z\in \mc{Z}}|\hat{g}(z)-g(z)|=O(\omega_n), \hspace{0.3in}
\sup_{z\in \mc{Z}}|\hat{v}_1(z)-v_1(z)|=O(\omega_n), \hspace{0.2in} w.p.1,
\eeq
where $\tau_m=h_1^{p+1}+\sqrt{\log(1/h_1)/(m h_1)}$ and
$\omega_n=h_2^{p+1}+ \sqrt{\log(1/h_2)/(n h_2)}$ as defined in Theorem 2.
\end{lemma}
\noindent {\it Proof of Lemma 2.} \quad It is sufficient to show (\ref{lem2.1}).
The strong uniform convergence rate $\tau_m$
for $\hat{f}$ was obtained by Horng (2006), which is based on the arguments in
Silverman (1978) and Mack and Silverman (1982) and the equivalent kernel representation, we follow the similar argument used in the proof of Lemma 1. Recall that
$v^\ast_{i,x}=\{x_i-f(z_{i,x})\}^2$, and  $v_{i,x}=\{x_i-\hat{f}(z_{i,x})\}^2=
v^\ast_{i,x}+2\{x_i-f(z_{i,x}\}\{\hat{f}\{z_{i,x}-f(z_{i,x})\}+\{\hat{f}(z_{i,x})-f(z_{i,x})\}^2$.
Applying a local polynomial fit to $(z_\ix, v_\ix)$, $i=1, \ldots, m$,
the second and third terms of the resulting estimator tend to 0 with probability 1,
and the leading term has the strong uniform convergence rate $\tau_m$
by using the same argument for $\hat{f}$.

\noindent {\it Proof of Theorem 2.} \quad The proof follows Lemma 2 and the uniform
version of Slutsky's Theorem. It is only needed to note that, if (A2$^\dag$) and
(A4$^\dag$) hold, $A_N=\Phi(f, g, v_1, v_2)$ has bounded partial derivative in each
argument, and thus satisfies Lipschitz continuity.

\noindent {\it Proof of Theorem 3.} \quad For a given $Z=z$, one can see that
``hypothetical'' estimator $A_M(z)$ is in fact a two-sample U-statistic. The
argument used in the theory of U-statistics can be applied here. The unbiasedness of $A_M(z)$
is obvious. For the asymptotic variance at a given $z$, put
$h(X, Y; z)=1_\ps(Y-X|Z=z)-A(z)$, $h^\ast_{0,0}=E\{h(X, Y; z)\}\equiv 0$.
$h^\ast_{1,0}(X; z)=E\{h(X, Y; z)|X\}$,
$h^\ast_{0,1}(Y; z)=E\{h(X, Y; z)|Y\}$.
Note that
\bea
h^\ast_{0,1}(Y; z)&=&P(Y\geq X|Y, Z=z)\\
&=&P\left(f(z)+\epsilon_1\sqrt{v_1(z)}\leq g(z)+\epsilon_2 \sqrt{v_2(z)}\,\Big|\,
\epsilon_2\right)\\
&=&P\left(\epsilon_1\leq \sqrt{\frac{v_2(z)}{v_1(z)}} \epsilon_2+
\frac{g(z)-f(z)}{\sqrt{v_1(z)}}\, \Big|\, \epsilon_2 \right)\equiv h_{1,0}(\epsilon_2; z),
\eea
and similarly $h^\ast_{1,0}(X; z)\equiv h_{1,0}(\epsilon_1; z)$, i.e.,
$\xi_{1,0}^2\equiv \mx{var}\{h_{1,0}^\ast(Y; z)\}$,
$\xi_{0,1}^2\equiv \mx{var}\{h_{0,1}^\ast(Y; z)\}$ as specified in Theorem 3.
The unbiasedness of $A_M(z)$ is obvious from $h_{0, 0}^\ast\equiv 0$. For
the variance calculation, after some counting techniques, one has,
\beq \label{var-auc-m}
\mx{var}\{A_M(z)\}=\frac{1}{mn} \sum_{c=0, 1}\sum_{d=0,1}
C_c^1 C_{1-c}^{m-1} C_d^1 C_{1-d}^{n-1} \xi_{c,d}
=\frac{\xi_{1,0}^2(z)}{m}+\frac{\xi_{0,1}^2(z)}{n}+o\left(\frac{1}{m+n}\right),
\eeq
where $C_k^n$ is the combination of choosing $k$ from $n$. This proves (\ref{thm3.1}).

To show the asymptotic normality (\ref{thm3.2}), define
$$T_{m,n}(z)=\sqrt{m+n} \left\{\frac{1}{m} \sum_{i=1}^m h_{1,0}^\ast(x_{i,z})+
\frac{1}{n} \sum_{j=1}^n h_{0,j}^\ast(y_\jz)\right\},$$ which is in fact the projection of
$\sqrt{m+n}\{A_M(z)-A(z)\}$ on the space formed by random variables of the form of
$\{\sum_{i=1}^m \psi(x_\iz)+\sum_{j=1}^n \psi^\ast(y_\iz)\}$, where $\psi$ and
$\psi^\ast$ are arbitrary measurable functions.
From H\'ajek's Projection Theorem and (\ref{var-auc-m}), we have,
as $m, n\rightarrow \infty$,
$$\mx{var}\{\sqrt{m+n}A_M(z)-T_{m,n}(z)\}=
\mx{var}\{\sqrt{m+n}A_M(z)\}-\mx{var}\{T_{m,n}(z)\}\longrightarrow 0,$$
which, together with unbiasedness, implies that $\sqrt{m+n}\{A_M(z)-A(z)\}$ is asymptotically equivalent to
$T_{m,n}(z)$.
Then following central limit theorem, when $n/(m+n)\rightarrow \lambda^\ast$
and $\min\{\xi_{1,0}^2(z), \xi_{0,1}^2(z)\}>0$, $T_{m,n}(z)$ has the limiting distribution
as specified in (\ref{thm3.2}). So does $\sqrt{m+n}\{A_M(z)-A(z)\}$.

\noindent {\it Proof of Theorem 4.}\quad Define
$w_{ij}=y_\iz-x_\iz$ and $\hat{w}_{ij}=\hat{y}_\iz-\hat{x}_\iz$, and
the dependences of $w_{ij}$ and $\hat{w}_{ij}$ on $x_\iz, y_\jz, z_\ix,
z_\jy$ and $z$ are suppressed for simplicity.
Let $a_1(z)=g(z)-f(z)$, $a_2(z_\jy, z)=\sqrt{v_2(z)/v_2(z_\jy, z)}$,
$a_3(z_\ix, z)=-\sqrt{v_1(z)/v_1(z_\ix)}$, $a_4(z_\jy, z)=-g(z_\jy)a_2(z_\jy, z)$,
$a_5(z_\ix, z)=-f(z_\ix)a_3(z_\ix, z)$, and then
\bea
w_{ij}&=&a_1(z)+a_2(z_\jy, z) y_j+a_3(z_\ix,z)x_i+a_4(z_\jy,z)+a_5(z_\ix,z),\\
\hat{w}_{ij}&=&\hat{a}_1(z)+\hat{a}_2(z_\jy, z) y_j+\hat{a}_3(z_\ix,z)x_i
+\hat{a}_4(z_\jy,z)+\hat{a}_5(z_\ix,z),
\eea
where ``$\hat{\quad}$'' is the generic notation for estimated quantities.
By analogy to the proof of  Lemma 2 with the assumptions (A3$^\dag$) and
(A5$^\dag$) replaced by (A3$^\ast$) and (A5$^\ast$), we obtain
weak (in probability) uniform consistency of $\hat{f}$,  $\hat{g}$, $\hat{v}_1$ and
$\hat{v}_2$. This is sufficient for our purpose, the reason of which will be singled
out below. Again by analogy to the proof of Theorem 2 with the uniform version of
Slutsky's Theorem (in probability instead of almost sure), we have, for a given $z$,
$\hat{a}_1(z)\stackrel{p}{\rightarrow}a_1(z)$, $\sup_{z_\jy} |\hat{a}_k(z_\jy, z)-
a_k(z_\jy,z)|=o_p(1)$, $\sup_{z_\ix}|\hat{a}_l(z_\ix, z)-a_l(z_\ix,z)|=o_p(1)$,
for $k=2, 4$ and $l=3, 5$. Since $\epsilon_{1,i} \stackrel{\mx{i.i.d.}}{\sim} F^\ast$,
one has $\epsilon_{1,i}=O_p(1)$and, analogously, $\epsilon_{2, j}=O_p(1)$,
regardless of $i$ and $j$.
Also note that $f$, $g$, $v_1$ and $v_2$ are bounded on $\mc{Z}$, then we obtain
$\sup_{i, j, z_\ix, z_\jy}|\hat{w}_{ij}-w_{ij}|=o_p(1)$ that only depends on the given $z$.

To show (\ref{thm4.1}), we observe that
$E[\{\wh{A}_M(z)-A_M(z)\}^2]=E_{0,0}+E_{1,0}+E_{0,1}+E_{1,1}$,
where
\bea
E_{0,0}=\frac{1}{m^2 n^2} \sum_{i\neq i', j\neq j'} E \Big[\{1_\ps(\hat{w}_{ij})-1_\ps(w_{ij})\}
\{1_\ps(\hat{w}_{i'j'})-1_\ps(w_{i'j'})\}\Big],
\eea
while $E_{1,0}$, $E_{0,1}$ and $E_{1,1}$ are defined in the same way, with $E_{1,0}$
corresponds to $\sum_{i=i', j\neq j'}$, $E_{0, 1}$ to $\sum_{i\neq i', j=j'}$
and $E_{1,1}$ to $\sum_{i=i', j=j'}$. We first focus on $E_{0,0}$,
\be \label{E00}
E_{0,0}&=&\frac{1}{m^2 n^2} \sum_{i\neq i', j\neq j'}
\Big\{P(\hat{w}_{ij}\geq 0, \hat{w}_{i'j'}\geq 0)+P(w_{ij}\geq 0, {w}_{i'j'}\geq 0)\nonumber \\
&&\hspace{1in} -P(\hat{w}_{ij}\geq 0, {w}_{i'j'}\geq 0)
-P(w_{ij}\geq 0, \hat{w}_{i'j'}\geq 0) \Big\} \nonumber \\
&\leq& \sup_{i, i', j, j'}\Big|
P(\hat{w}_{ij}\geq 0, \hat{w}_{i'j'}\geq 0)+P(w_{ij}\geq 0, {w}_{i'j'}\geq 0) \nonumber \\
&&\hspace{0.5in} -P(\hat{w}_{ij}\geq 0, {w}_{i'j'}\geq 0)
-P(w_{ij}\geq 0, \hat{w}_{i'j'}\geq 0)\Big|.
\ee
For any given $z$, from Slutsky's Theorem, we have
$(\hat{w}_{ij}, \hat{w}_{i'j'})^T$, $(\hat{w}_{ij}, {w}_{i'j'})^T$ and
$({w}_{ij}, \hat{w}_{i'j'})^T$ converge in probability to $(w_{ij}, w_{i'j'})^T$
uniformly in all arguments except $z$, which implies uniform convergence
in distribution. Therefore the four sequences of probabilities in (\ref{E00})
all uniformly converge to $P(w_{ij}\geq 0, {w}_{i'j'}\geq 0)$ as $m, n\rightarrow \infty$,
which leads to $E_{0,0}\rightarrow 0$. From the above argument, one can see that
the weak uniform consistency is sufficient, also that the convergence
rates cannot be preserved for evaluating upper bounds for those probability differences.
Using similar arguments, it is easy to show that $E_{1,0}=O(E_{0,0}/m)$,
$E_{0,1}=O(E_{0,0}/n)$ and $E_{1,1}=O\{E_{0,0}/(mn)\}$. This completes the proof of
Theorem 4.
\newpage

\nocite{*}
\bibliography{ref}

%\label{simres} \caption{{\em Simulation results for two scenarios with normal noise:
%(a) mean functions non-diseased (lower curve) and diseased (upper curve); (b)
%variance functions for non-diseased (lower curve) and diseased (upper curve); (c)
%True AUC curve along with nonparametric estimates with and without
%the normal error assumption and the confidence bands obtained using
%the two bootstrap methods.}}

\begin{figure}[htbp]
\begin{center}
{\includegraphics[height=1.6in,width=2.7in,angle=270]{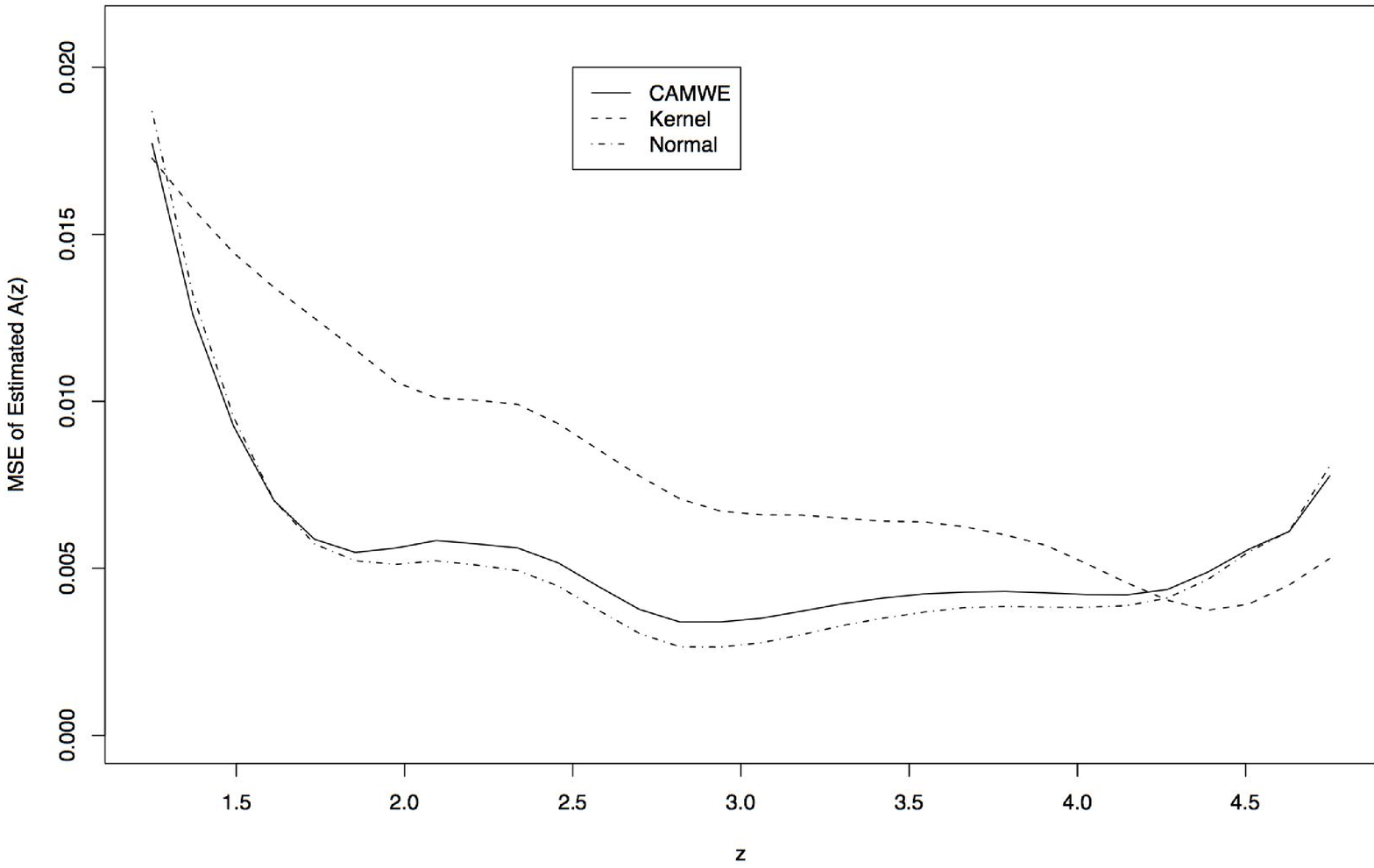}}
\hspace{-0.1in}
{\includegraphics[height=1.6in,width=2.7in,angle=270]{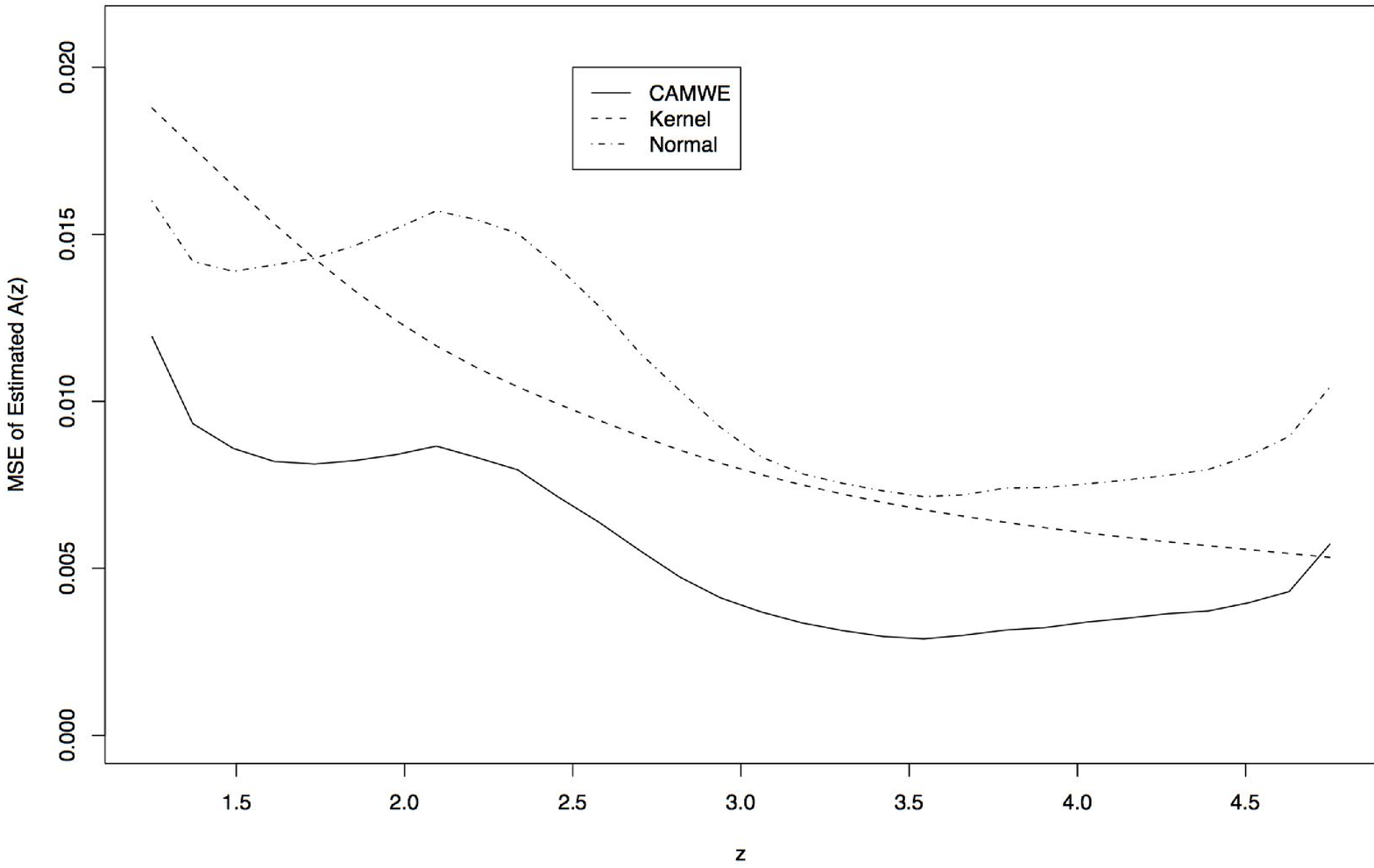}}
\hspace{-0.1in}
{\includegraphics[height=1.6in,width=2.7in,angle=270]{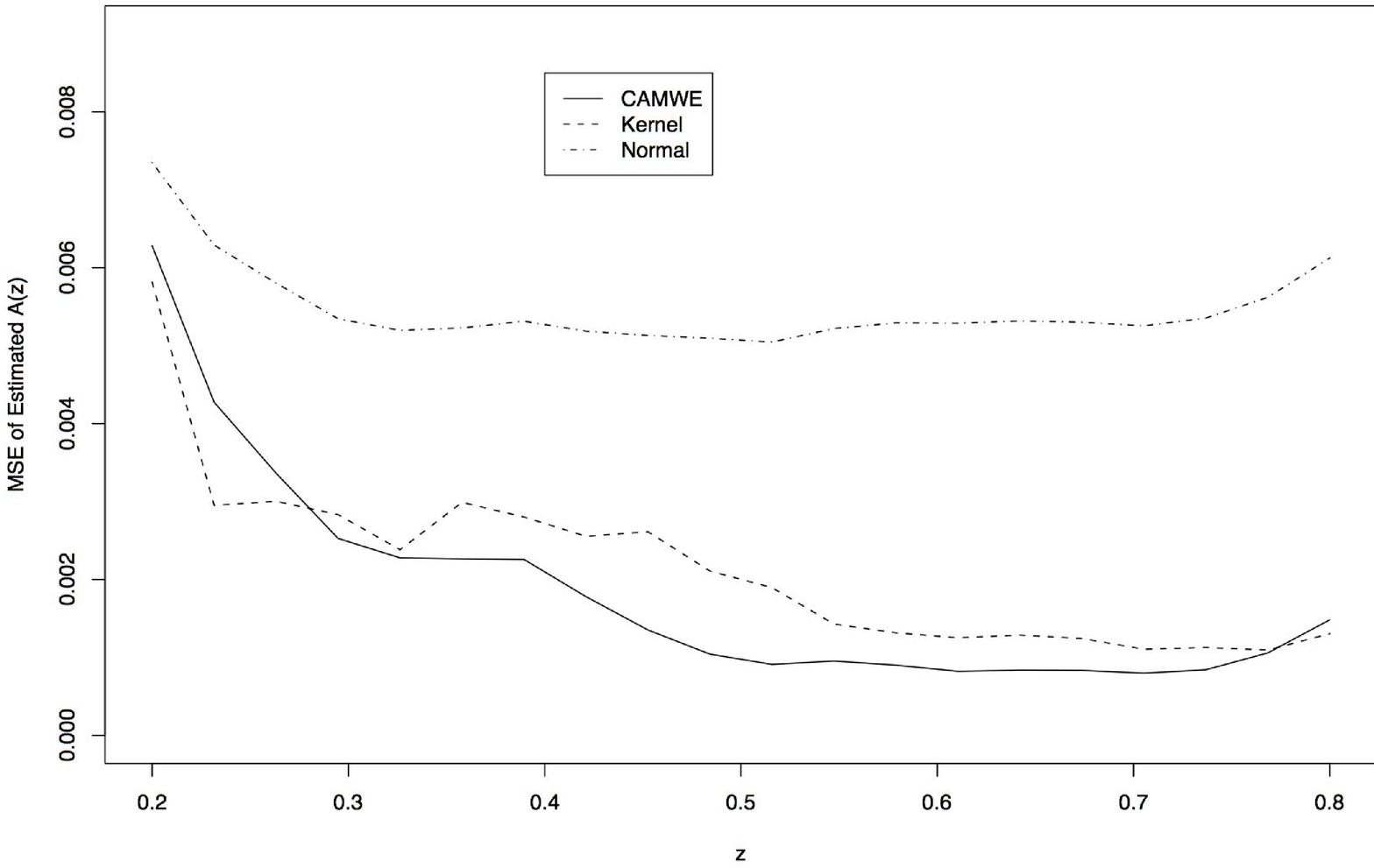}}
\vspace{-0.3in}

{\includegraphics[height=1.6in,width=2.7in,angle=270]{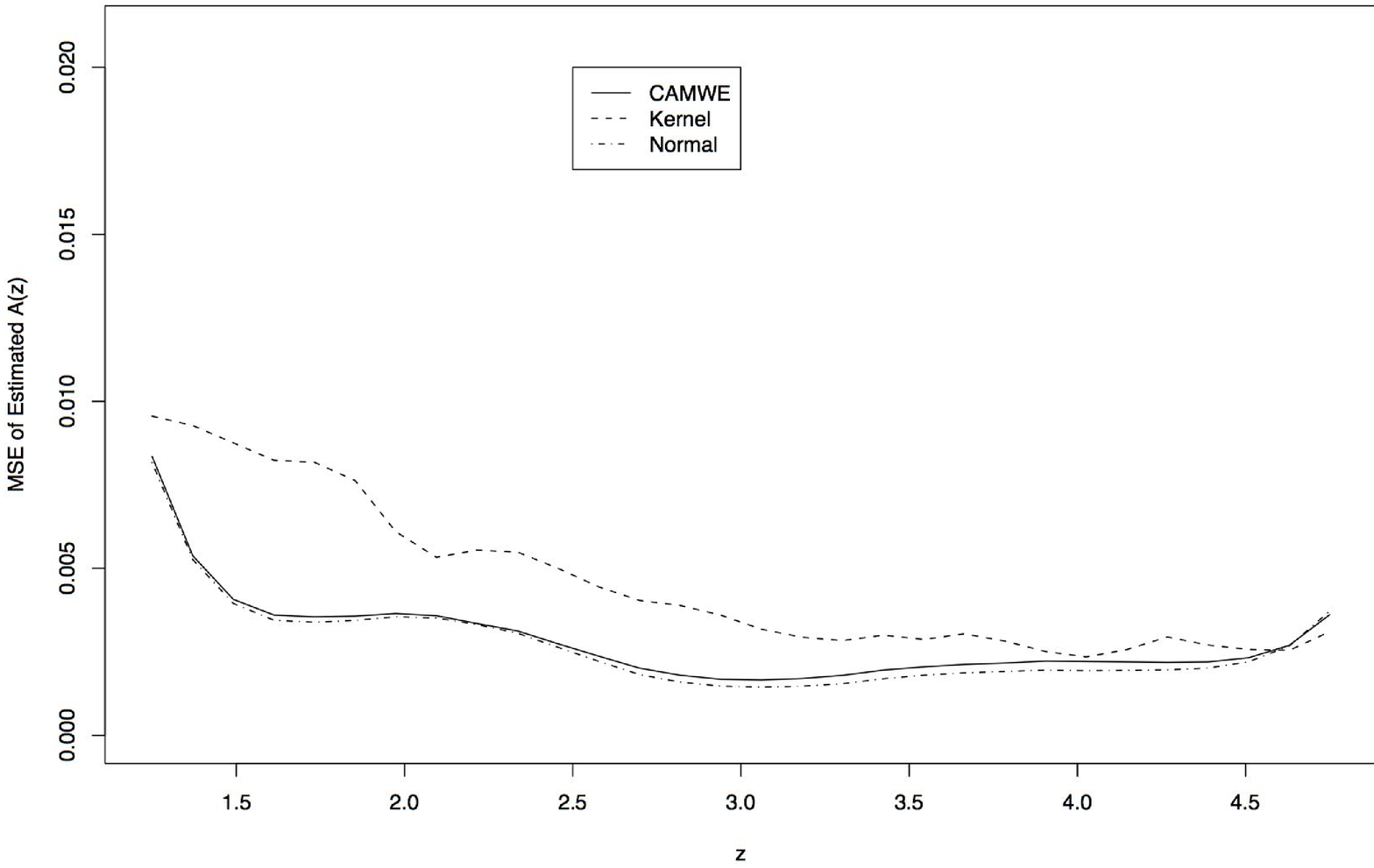}}
\hspace{-0.1in}
{\includegraphics[height=1.6in,width=2.7in,angle=270]{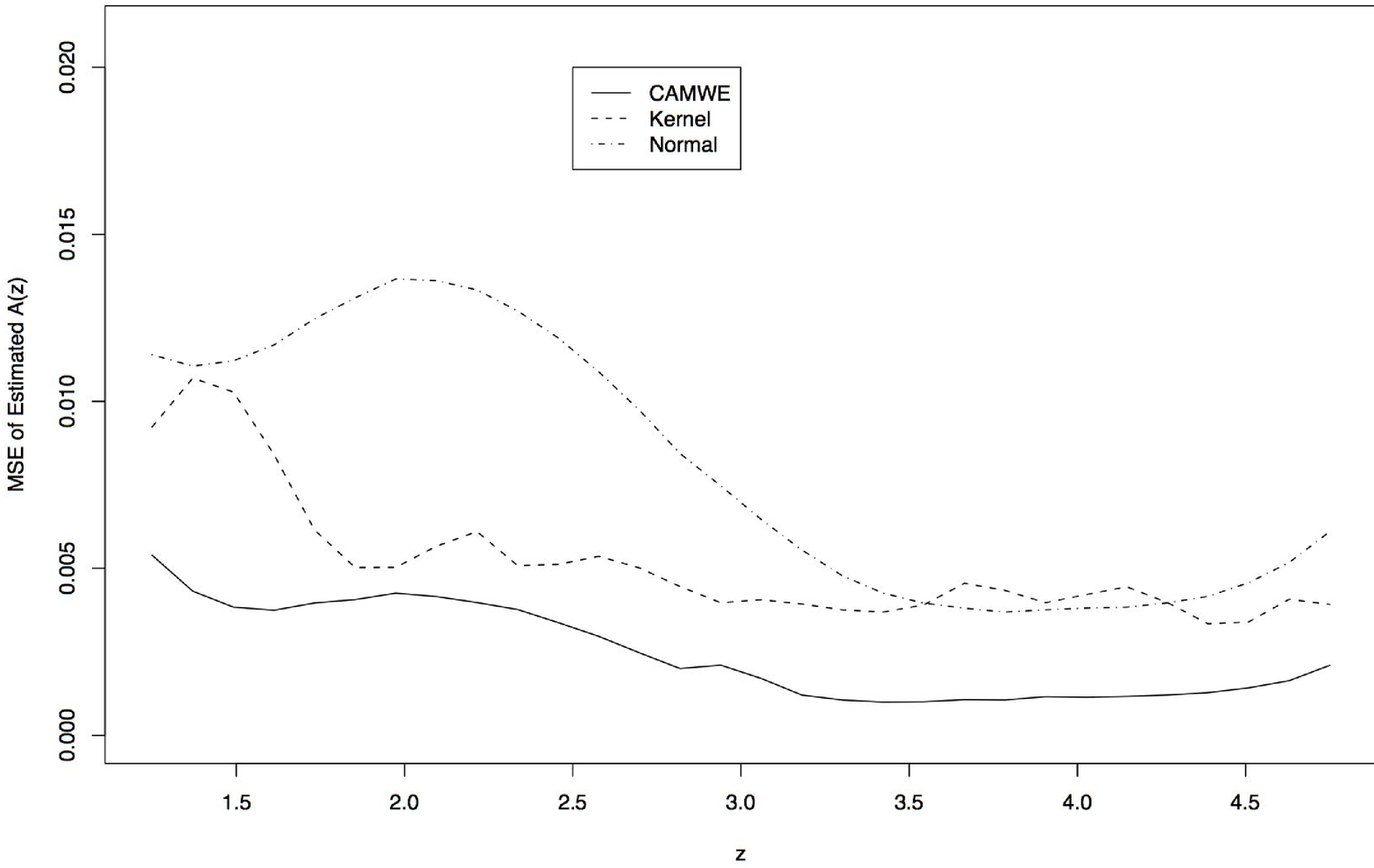}}
\hspace{-0.1in}
{\includegraphics[height=1.6in,width=2.7in,angle=270]{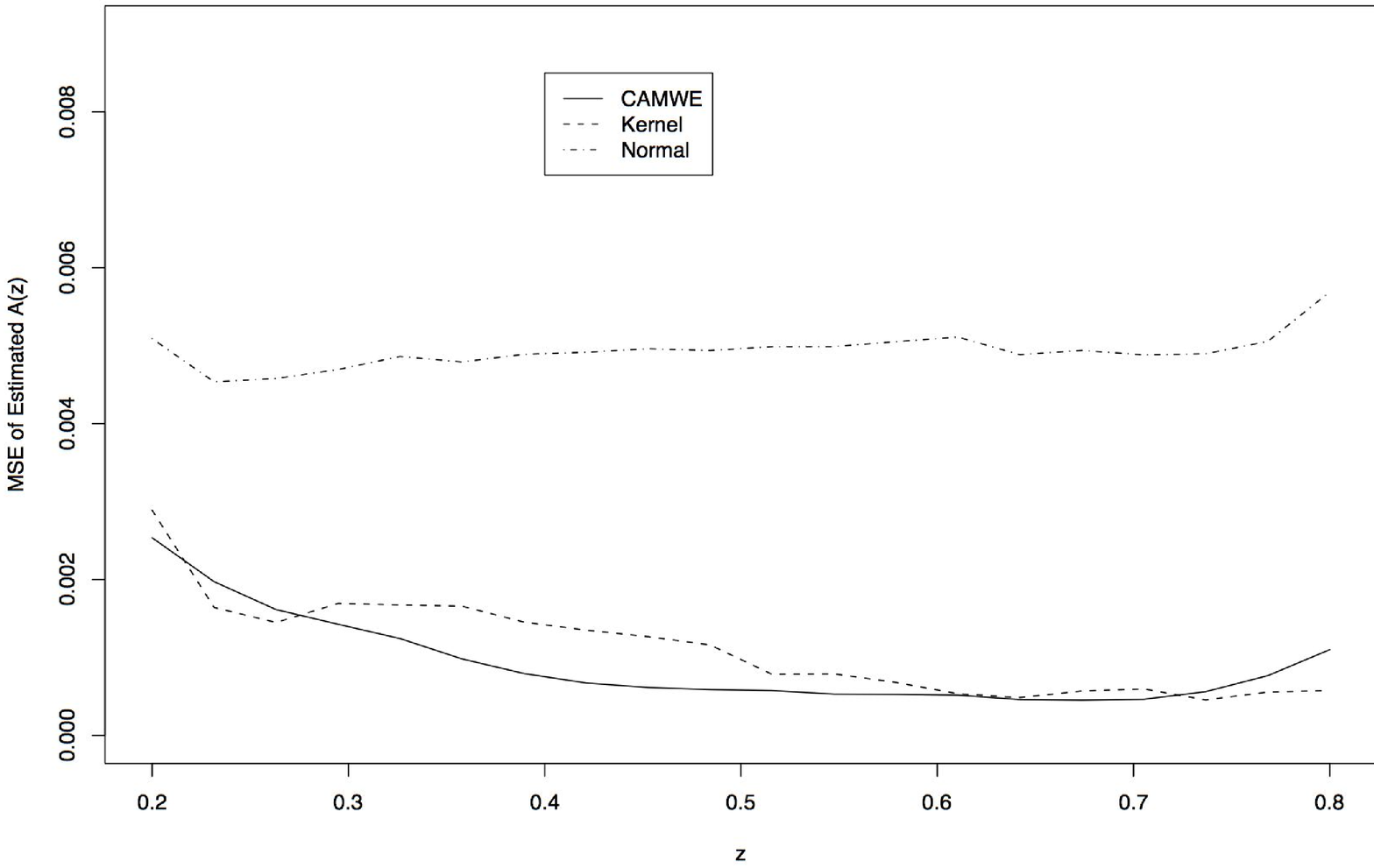}}

\vspace{-0.1in}
\caption{\label{simu-mse} {Top Row: Simulation results for the three models with normal (left),  Student-$t$ with 3 degrees of freedom (middle) and log-normal errors (right). Shown are Monte Carlo averages of Mean Squared Errors (MSE) of three estimators, $\wh{A}_M$ (CAMWE, solid), $\wh{A}_N$ (Normal, dash-dotted) and $\wh{A}_K$ (Kernel, dashed) at different values of $z$ with moderate sample sizes $n=m=40$.
Bottom Row: The  simulation results in the same scenarios with larger sample sizes $n=m=100$.}}
\end{center}
\end{figure}

\begin{figure}[htbp]
\begin{center}
{\includegraphics[height=1.6in,width=2.7in,angle=270]{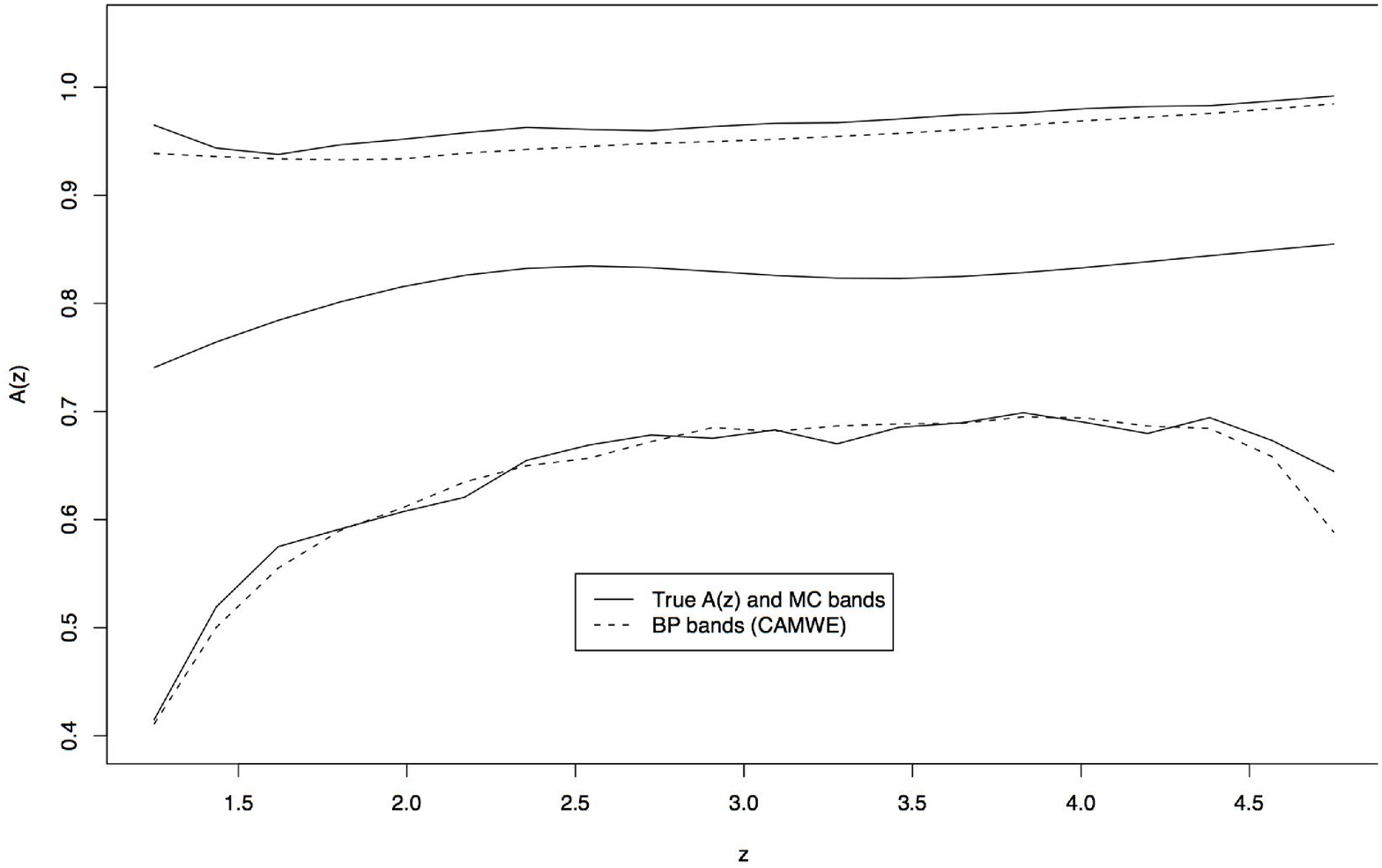}}
\hspace{-0.1in}
{\includegraphics[height=1.6in,width=2.7in,angle=270]{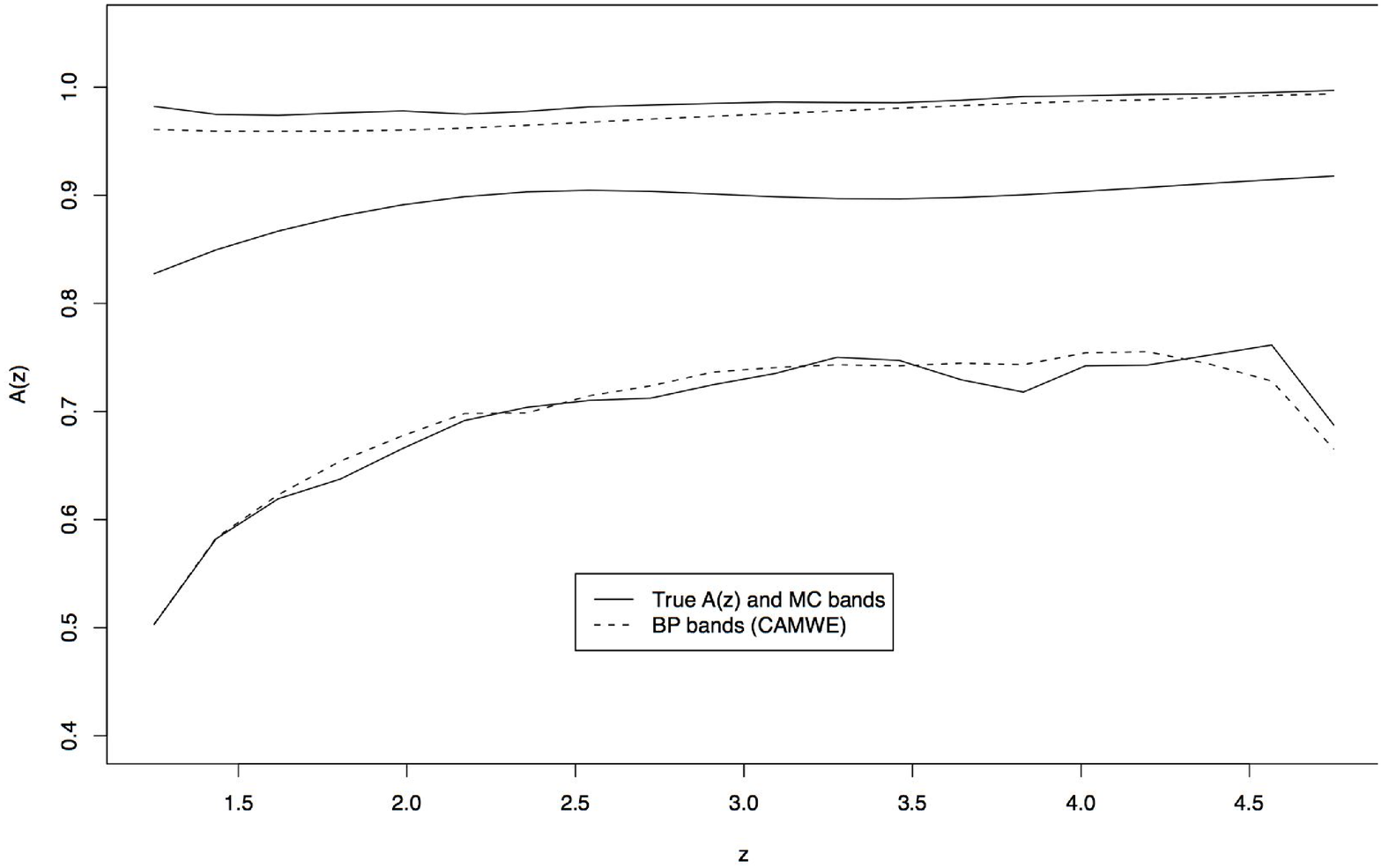}}
\hspace{-0.1in}
{\includegraphics[height=1.6in,width=2.7in,angle=270]{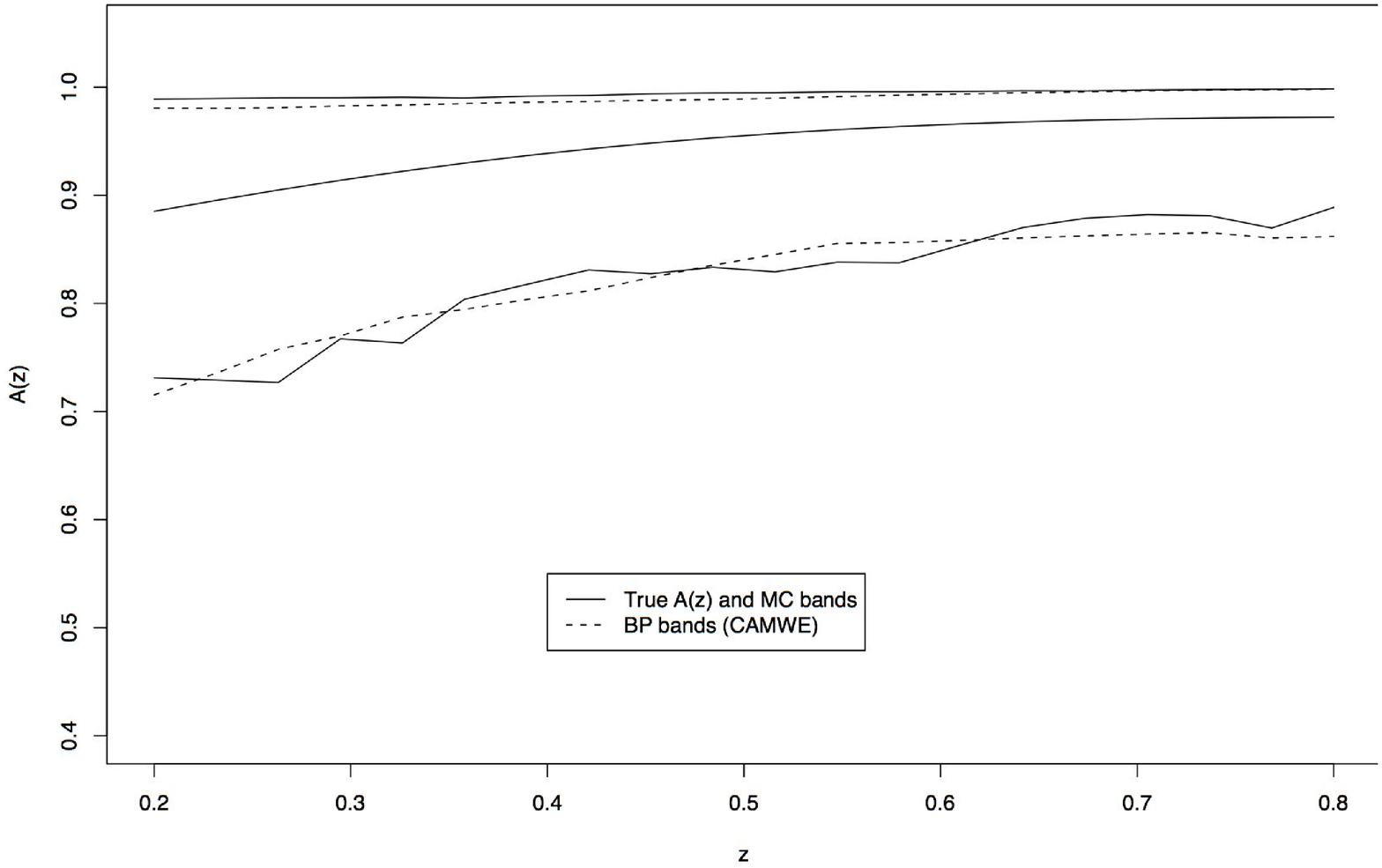}}
\vspace{-0.3in}

{\includegraphics[height=1.6in,width=2.7in,angle=270]{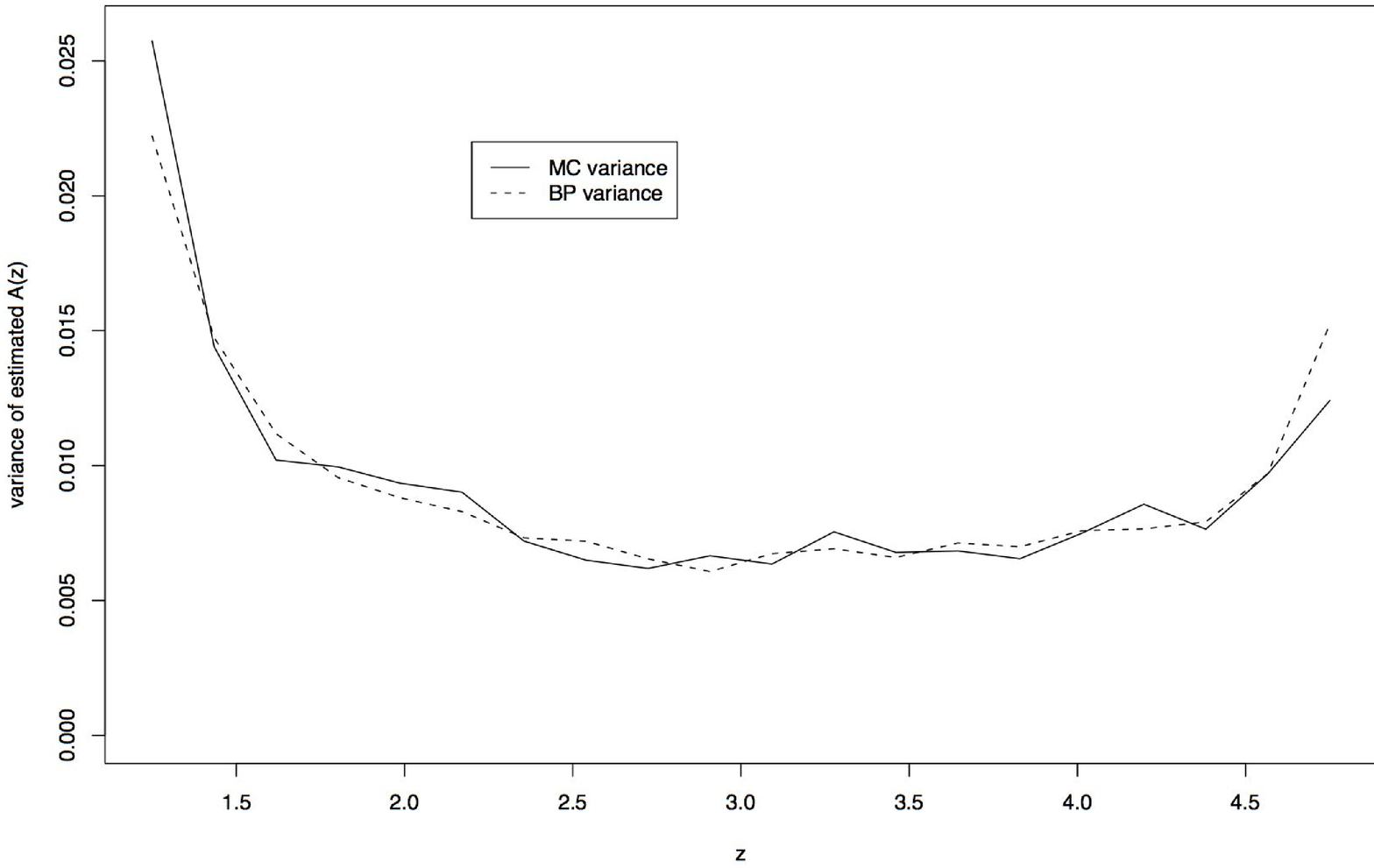}}
\hspace{-0.1in}
{\includegraphics[height=1.6in,width=2.7in,angle=270]{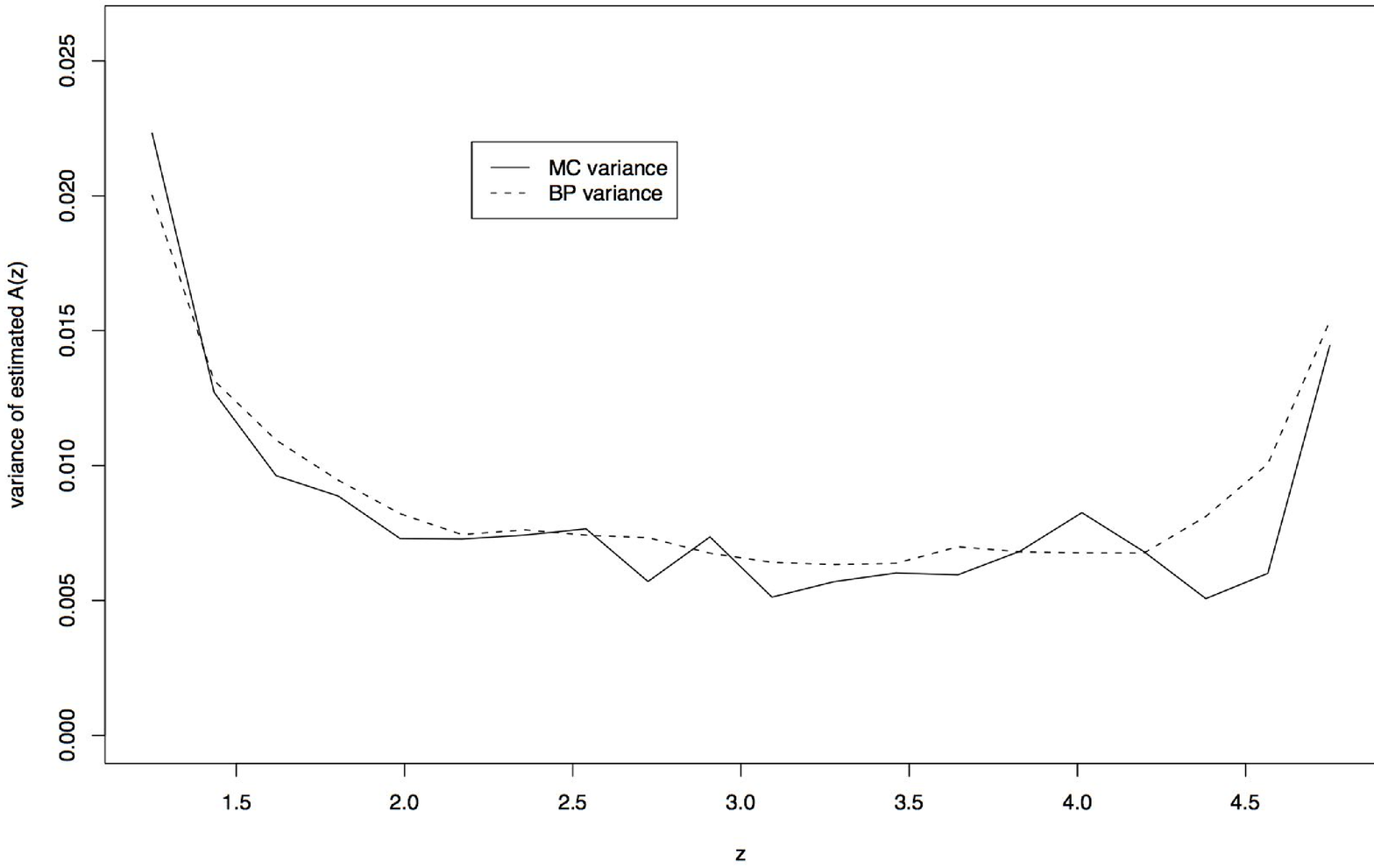}}
\hspace{-0.1in}
{\includegraphics[height=1.6in,width=2.7in,angle=270]{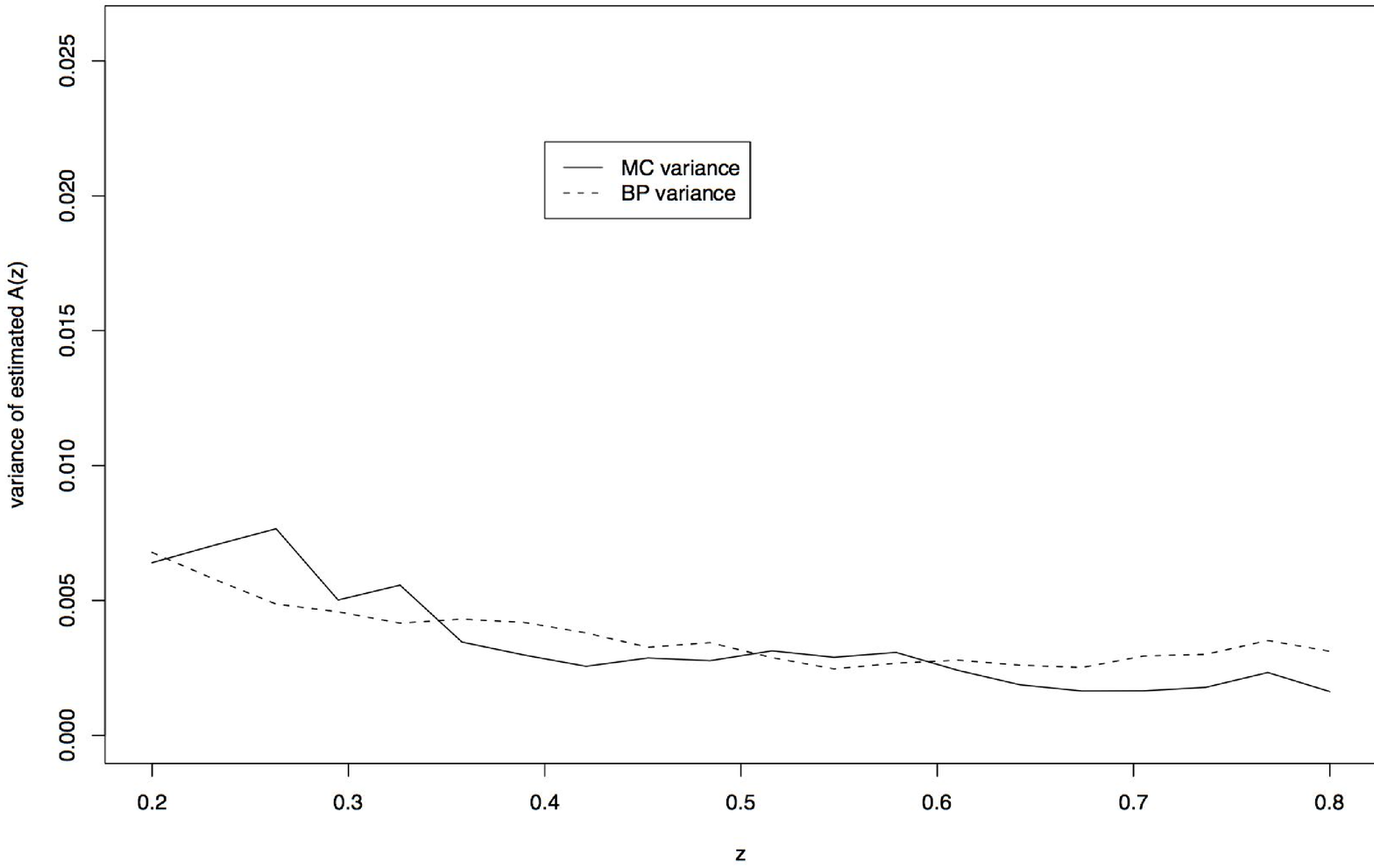}}
\vspace{-0.1in}

\caption{\label{simu-bc} {Simulation results of 95\% pointwise bootstrap confidence bands (top row) and variance comparisons (bottom row) for three models  with normal (left), Student-$t$ with 3 degree of freedom (middle)  and log-normal (right) noise with the same settings as in Figure \ref{simu-mse} and sample sizes $n=m=40$. Top row: True AUC and 95\% pointwise Monte Carlo bands (solid) obtained from 500 runs, and the Monte Carlo averages of 95\% pointwise bootstrap bands (dashed). Bottom row: Monte Carlo variance estimates (solid) obtained from 500 runs, and the Monte Carlo averages of bootstrap variance estimates (dashed), as described  in Section 4.1. }}
\end{center}
\end{figure}

\begin{figure}[htbp]
\begin{center}
{\includegraphics[height=3.4in,width=3in,angle=270]{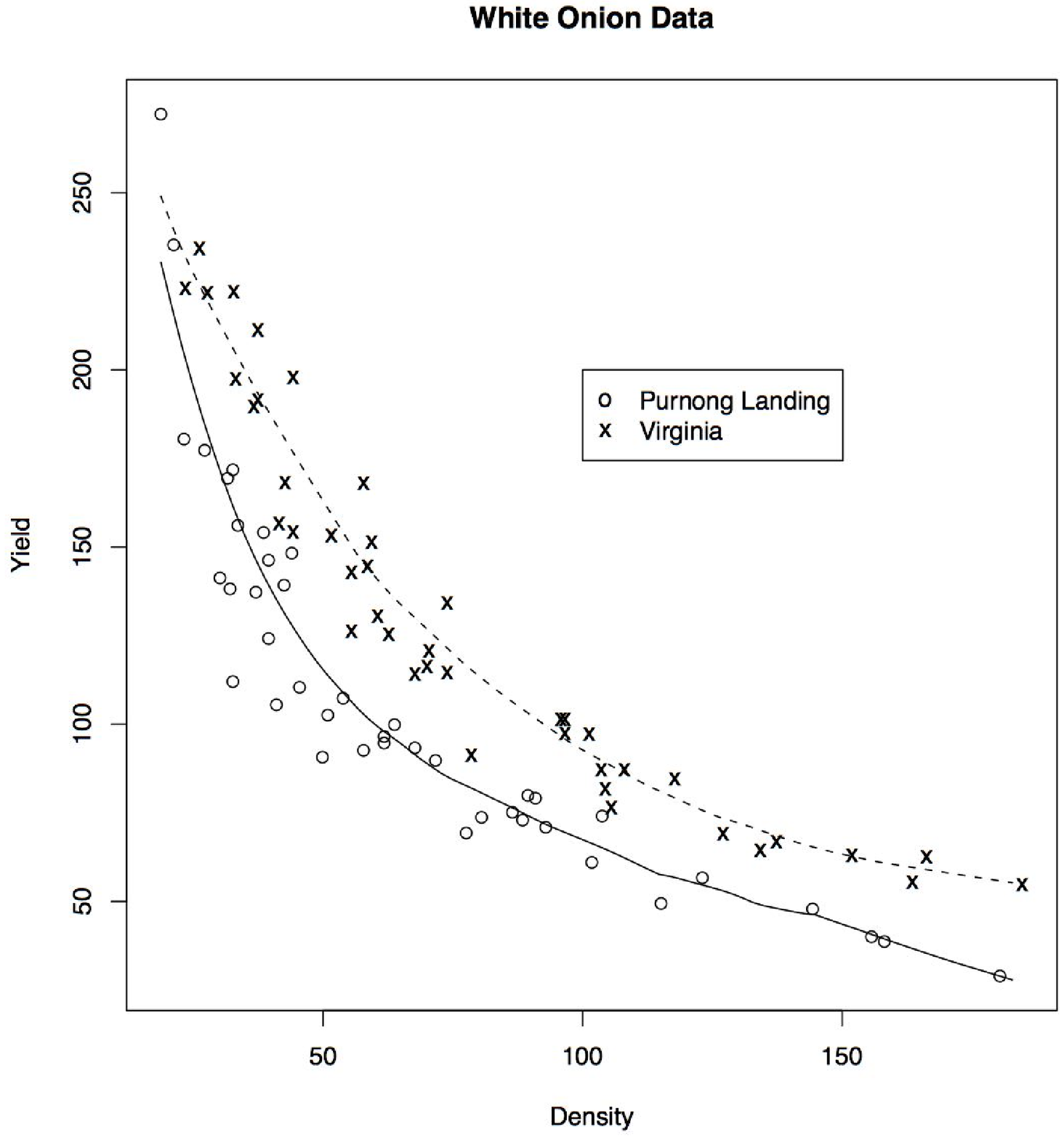}}
\hspace{-0.5in}
{\includegraphics[height=3.4in,width=3in,angle=270]{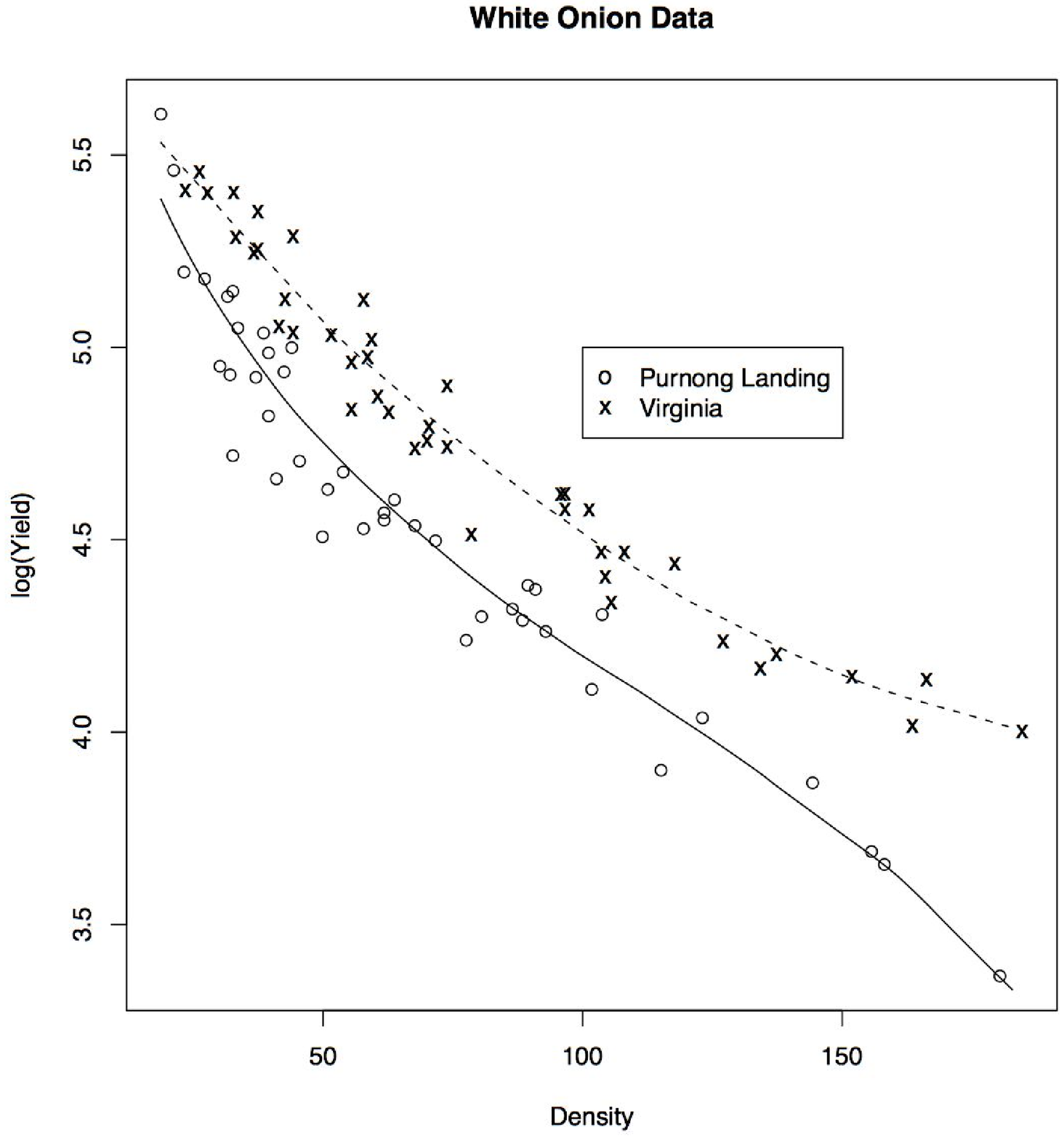}}
\caption{\label{fig:wo} { Spanish Onion Data with response on the orginal scale (top)  and the logarithmic scale (bottom), with the smooth estimates of the mean functions for two populations, Pumong Landing (solid) and Virginia (dashed). }}
\end{center}
\end{figure}

\begin{figure}
\begin{center}
{\includegraphics[height=6in,width=3in,angle=270]{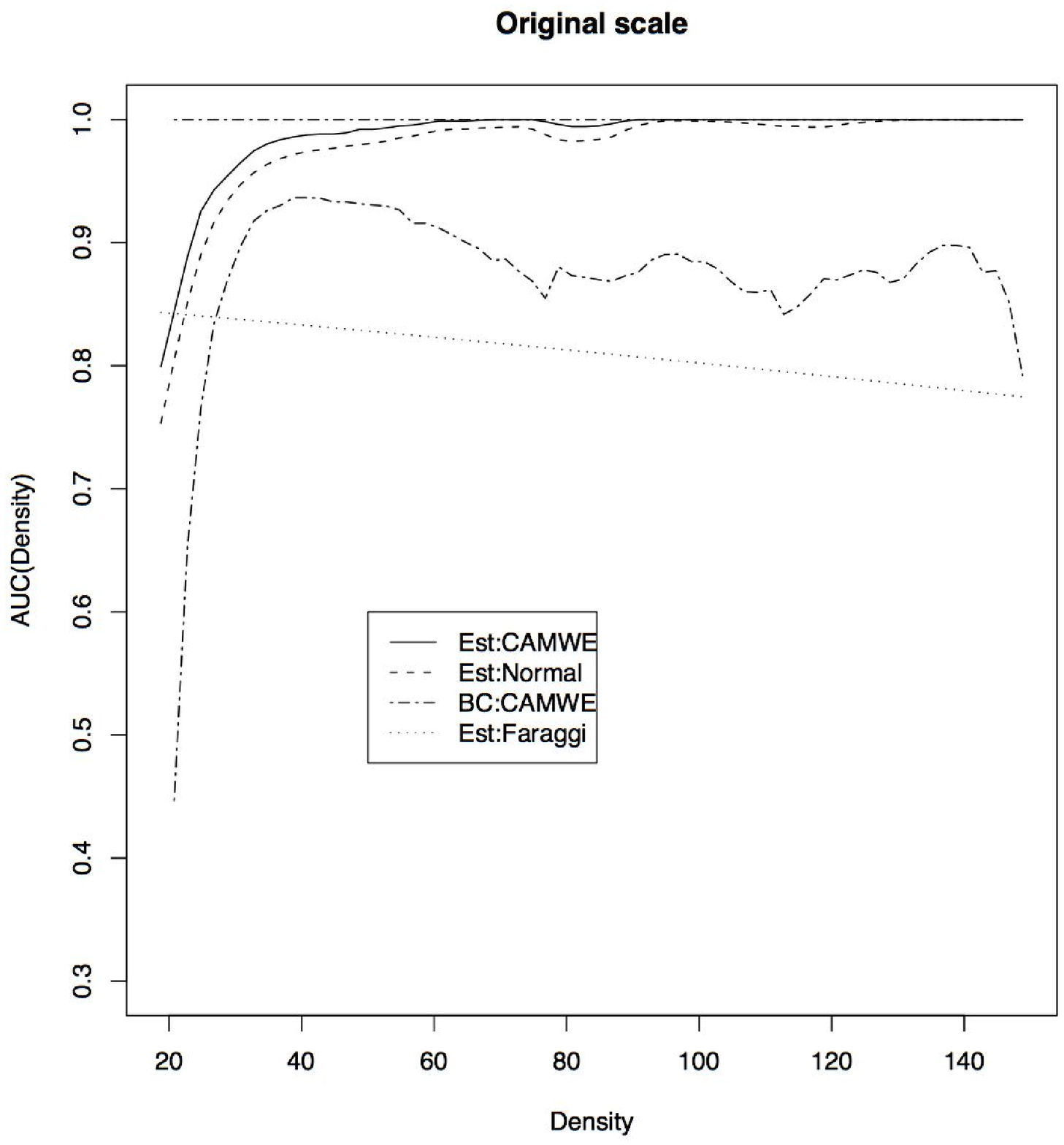}} 
{\includegraphics[height=6in,width=3in,angle=270]{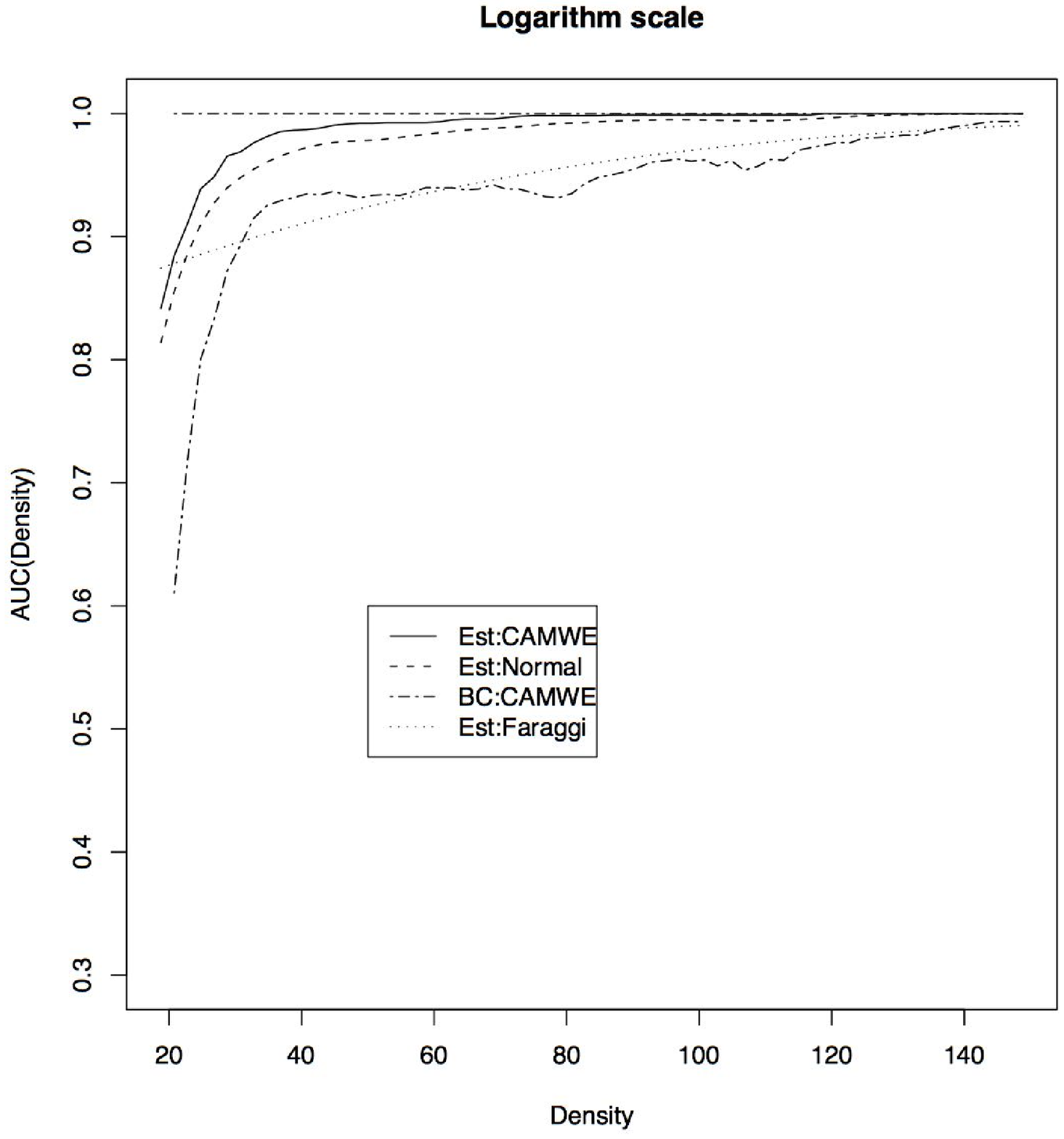}}
\caption{\label{fig:woauc} {Top panel: Comparison of estimated functional relationship
between AUC and density obtained using the nonparametric approach
with and without normal noise assumption, denoted by Normal and CAMWE respectively, with the parametric estimate following Faraggi (2003). Also shown are the 95\% pointwise confidence bands obtained from nonparametric Bootstrap method. Bottom panel: Same comparison as in the top panel with response on the logarithmic scale. }}
\end{center}
\end{figure}

\end{document}